\documentclass[structabstract]{aa}  
\usepackage{graphicx}
\usepackage{txfonts}
\usepackage{natbib}
\usepackage[draft]{hyperref} 
\hypersetup{
    colorlinks, linkcolor={Black},
    citecolor={Black}, urlcolor={RoyalBlue}
}
\usepackage[usenames,dvipsnames]{xcolor}
\renewcommand{\arcsec}{$^{\prime\prime}$}
\renewcommand{\arcmin}{$^{\prime}$}  
\newcommand{\kms}{\nobreak{km~s$^{-1}$}} 
\newcommand{\mi}{$\,\mu$m} 
\newcommand{\halpha}{H$\alpha$} 
\newcommand{\hbeta}{H$\beta$} 

\newcommand{\Msun}{$\,M_{\sun}$}
\newcommand{\Zsun}{$\,Z_{\sun}$}

\newcommand{\sfrprime}{$SFR^{\prime}$}
\newcommand{\taubf}{$\tau_{B}^{f}$}
\newcommand{\condon}{$\rm S_{T}~(0.5\,cm)$}
\newcommand{\fcov}{$f_{\rm cov}$} 
\newcommand{\fesc}{$f_{\rm esc}$} 
\newcommand{\mcl}{$M_{\rm cl}$}
\newcommand{\po}{$\log(p_{0}/k)$}
\newcommand{\Comp}{$\log(\mathcal{C})$}
\newcommand{\Age}{$t$} %{$\mathcal{T}$}

\newcommand{\SFR}[3]{$SFR^{#1}_{#2}=#3\,M_{\sun}\,$yr$^{-1}$}
\newcommand{\hs}[1]{$h_{\rm s} = #1\,$pc}
\newcommand{\RA}[3]{#1h#2m#3s}
\newcommand{\DEC}[4]{#1$^{\circ}$#2$^{\prime}$#3.$^{\prime\prime}$#4}
\newcommand{\IRAC}[1]{IRAC~#1$\,\mu$m}
\newcommand{\MIPS}[1]{MIPS~#1$\,\mu$m}

\newcommand{\SPIRE}[1]{SPIRE~#1$\,\mu$m}

\begin{document}
   \title{The dust SED of dwarf galaxies}

   \subtitle{I. The case of NGC~4214}

   \author{I. Hermelo\inst{1}
          \and
          U. Lisenfeld\inst{1}
          \and
          M.  Rela\~no\inst{1}
          \and 
          R. J. Tuffs\inst{2}
          \and
          C. C. Popescu\inst{3}
          \and
          B. Groves\inst{4}
          }

\institute{
Departamento de F\'isica Te\'orica y del Cosmos, Universidad de Granada, Spain\\
\email{israelhermelo@ugr.es, ute@ugr.es, mrelano@ugr.es}
\and
Max-Planck-Institut f\"ur Kernphysik, Saupfercheckweg 1, 69117 Heidelberg, Germany
\and
Jeremiah Horrocks Institute for Astrophysics and Supercomputing, University of  Central Lancashire, PR1 2HE, Preston, U.K.
\and
Max-Planck-Institut f\"ur Astronomie, K\"onigstuhl 17, D-69117 Heidelberg,  Germany
}

%    \date{Received September 15, 1996; accepted March 16, 1997}

% \abstract{}{}{}{}{} 
% 5 {} token are mandatory
 
\abstract
{The dust properties in dwarf galaxies have been a matter of debate. High resolution data from SPITZER, HERSCHEL and PLANCK allow us to probe the entire spectral energy distribution (SED) of morphologically separated components of the dust emission from nearby galaxies and allow a more detailed comparison between data and models.}
{We wish to establish the physical origin of dust heating and emission based on radiation transfer models, which self-consistently connect the emission components from diffuse dust and the dust in massive star forming regions.}
{NGC~4214 is a nearby dwarf galaxy with a large set of ancillary data, ranging from the ultraviolet (UV) to radio, including maps from SPITZER, HERSCHEL and detections from PLANCK. We mapped this galaxy with MAMBO at 1.2\,mm at the IRAM 30\,m telescope. We extract separate dust emission components for the HII regions (plus their associated PDRs on pc scales) and for the diffuse dust (on kpc scales). We analyse the full UV to FIR/submm SED of the galaxy using a radiation transfer model which self-consistently treats the dust emission from diffuse and SF complexes components, considering the illumination of diffuse dust both by the distributed stellar populations, and by escaping light from the HII regions. While maintaining consistency with the framework of this model we additionally use a model that provides a detailed description of the dust emission from the HII regions and their surrounding PDRs on pc scales. Due to the large amount of available data and previous studies for NGC~4214 very few free parameters remained in the model fitting process.}
%results heading (mandatory)
{We achieve a satisfactory fit for the emission from HII+PDR regions on pc scales, with the exception of the emission at 8\mi, which is underpredicted by the model. For the diffuse emission we achieve a good fit if we assume that about 30-70\% of the emission escaping the HII+PDR regions is able to leave the galaxy without passing through a diffuse ISM, which is not an unlikely scenario for a dwarf galaxy which has recently undergone a nuclear starburst. We determine a dust-to-gas mass ratio of 350-390 which is close to the expected value based on the metallicity.}
{}


 \maketitle

%%%%%%%%%%%%%%%
%%%% INTRO %%%%
%%%%%%%%%%%%%%%

\section{Introduction}
\label{sec:intro}

Interstellar dust is a component of the interstellar medium (ISM) that is
present in all phases from dense molecular clouds to
the warm ionised regions around massive stars. It plays an important role
in star formation (SF) and in the overall energy
budget of a galaxy. Measurements of the dust re-emission from galaxies is a powerful technique to quantify the SF. The details however, are not fully understood yet
as the emission depends on several parameters as the geometry, the mixture between dust, stars and HII regions and also on the dust properties in the interstellar medium which are expected to vary as a function of metallicity and are known to depend on the environment. 

The study of dust properties (e.g. mass, extinction coefficient)
has been an object of debate. Much progress has
been achieved in recent years due to data  from  the satellites SPITZER,
HERSCHEL and PLANCK,
which allow for the first time to probe the entire wavelength range of
the spectral energy distribution (SED) of the dust emission,  from the mid-infrared (MIR) to the
submillimeter (submm), for a large number of galaxies at the best resolution ever.

The dust SED is a key observation for understanding the properties of dust.
The dust SED of dwarf galaxies frequently shows differences to those of spiral galaxies. The main differences are:
i) The SED of dwarf galaxies shows a relatively low emission at 8\mi, most likely due to a relatively lower content of polycyclic aromatic hydrocarbons (PAHs) at low metallicities (e.g. \citealt{2007ApJ...663..866D}, \citealt{2008ApJ...672..214G}, \citealt{2008ApJ...678..804E}).
%
% {\bf ii) The dust SED peaks at shorter wavelengths, which indicates hotter equilibrium grains and shows a steeper rising MIR continuum \citep[e.g.][]{1994A&A...285...51M}. This can be an indication for a lower relative contribution of emission from diffuse ("cirrus") dust (\citealt{1994A&A...285...51M}) and a higher contribution from very small grains (\citealt{2003A&A...407..159G}, \citeyear{2005A&A...434..867G}).}
%
ii) A submm "excess" has been found in the SED of many, mostly actively star-forming, low-metallicity galaxies (\citealt{2002A&A...382..860L},
% \citealt{2002ApJ...567..221P},
\citealt{2003A&A...407..159G}, \citeyear{2005A&A...434..867G}, \citealt{2006ApJ...652..283B}, \citealt{2009A&A...508..645G}, \citeyear{2011A&A...532A..56G}, \citealt{2010A&A...519A..67I}, \citealt{2010A&A...523A..20B}, \citealt{2012ApJ...745...95D}, \citealt{2011A&A...536A..17P}). Several reasons have been suggested to explain this excess: the existence of a large amount of cold ($\rm<10\,K$ ) dust (\citealt{2003A&A...407..159G}, \citeyear{2005A&A...434..867G}, \citealt{2009A&A...508..645G}, \citeyear{2011A&A...532A..56G}), different dust grain properties (
\citealt{2002A&A...382..860L},
\citealt{1995ApJ...451..188R},
\citealt{2007A&A...468..171M},
% \citealt{1994ApJ...427..155F},
% \citealt{1998ApJ...508..157D}
), or magnetic nanograins, which can produce magnetic dipole radiation at microwave and submillimeter wavelengths (\citealt{2012arXiv1205.6810D}).

It is also noteworthy that one sees a variety of FIR colours for gas rich dwarf galaxies, with examples of warm FIR SEDs (as found from IRAS colours - e.g. \citealt{1994A&A...285...51M}), or of cold FIR SEDs (as revealed by ISO and Herschel measurements extending longwards of 100\mi\ - e.g. \citealt{2002ApJ...567..221P}, \citealt{2010A&A...518L..52G}). 
Such variety may not be too surprising given that the star formation in dwarf galaxies is likely fundamentally bursty in nature. If we interpret the FIR SED in terms of the combination of cooler distributed ”cirrus” dust emission and warm dust emission from grains in SF regions, the latter opaque structures should be most prominent in the early stages of the evolution of a starburst leading to a warm FIR SED. over time the starburst fades, leaving the cirrus component and a cooler FIR SED.
Other factors potentially contributing to the observed variety in FIR colours are differences in the contribution of very small grains (e.g. \citealt{2003A&A...407..159G}, \citeyear{2005A&A...434..867G}) and the possible presence of extended cold dust outside the main star forming disk of the galaxy (\citealt{2002ApJ...567..221P}).

In order to interpret the dust SED of a galaxy and to understand the differences in the SEDs of dwarf galaxies, a physical model based on realistic dust properties and taking into account the heating and emission of dust immersed in a wide range of interstellar radiation fields (ISRFs) is needed. Ideally, radiation transport in a realistic geometry should be done, but this is often difficult due to the complex geometry and large number of parameters. Models can generally be classified into three broad groups: (1) modified blackbody fits, which are too simple to describe reality correctly but give a first idea of the dust temperature ranges, (2) semi-empirical models that try, in a simplified way, to describe dust immersed in a range of different ISRFs
(e.g. \citealt{2001ApJ...549..215D},
\citealt{2007ApJ...663..866D}, \citealt{2009A&A...508..645G},
\citeyear{2011A&A...532A..56G}, \citealt{2008MNRAS.388.1595D}, \citealt{2010ApJ...725..955N}) 
and (3) models that include full radiation transfer 
(e.g. \citealt{2011A&A...527A.109P} for spiral galaxies,
\citealt{2007A&A...461..445S} for star-burst galaxies,
see also \citealt{1998ApJ...509..103S}, 
\citealt{2000A&A...362..138P}, 
\citealt{2004A&A...414...45P},
\citealt{2001A&A...372..775M},
\citealt{2008A&A...490..461B},
\citealt{2010A&A...518L..39B},  \citeyear{2011ApJS..196...22B} and 
\citealt{2011ApJ...741....6M}),
which are the most precise description of a galaxy if all parameters, including the geometry,
are known.

The new IR facilities with their high angular resolution and sensitivity allow for the first time  a detailed comparison between models and data. For nearby, and thus spatially resolved, galaxies, the spatial variations of the dust SED can be probed and modelled. In particular, a good spatial resolution allows us to separate and treat the emission from dust heated by the UV radiation of massive stars in HII regions and their adjacent photodissociation regions (PDR), and the diffuse dust heated by the general ISRF.

We choose NGC~4214 as a test case because of the large amount of ancillary data and its proximity (1\arcsec$\sim$14\,pc, D=2.9\,Mpc; \citealt{2002AJ....123.1307M}), which allows us to observationally separate the dust emission from the two prominent central SF complexes and from the diffuse dust component in the disk. This allows us to apply a radiation transfer treatment constrained by measurements of direct light in UV to NIR and dust/PAH re-radiated light in both the diffuse and SF complexes.

We choose the radiation transfer model of \citet{2011A&A...527A.109P} for spiral galaxies since this self-consistently treats the dust emission from the diffuse and SF complex components, considering the illumination of diffuse dust both by the smoothly distributed stellar populations, and by escaping light from the HII regions in spiral disks. Although NGC~4214 is a  dwarf galaxy, rather than a spiral galaxy, it nevertheless exhibits the principle geometrical features of the \citet{2011A&A...527A.109P}  model. In particular, NGC~4214 shows an exponential disk distribution of the stellar light, the diffuse dust emission and the atomic gas emission, so is actually a good candidate to be fitted with a model for disk galaxies. While maintaining consistency with the framework of \citet{2011A&A...527A.109P} model, we use the model of Groves et al. (2008) to provide a detailed description of the dust emission from the HII regions and their surrounding PDRs on scales of tens of pc. These PDRs mark the transition from the ionised medium to the dense molecular gas left over from the highly opaque cloud out of which the stars formed. The model of \citet{2008ApJS..176..438G} takes into account the dynamical evolution of the distance of the PDRs from the central ionising star clusters due to the mechanical effect of stellar winds on the surrounding ISM. This model is therefore particularly suitable to describe the two prominent SF complexes in the centre of NGC~4214, since these show shell structures imaged by HST surrounding the two main star clusters of the galaxy, which may be delineating wind blown bubbles around the HII regions.

NGC~4214 is a Magellanic starbursting dwarf irregular galaxy (\citealt{1991S&T....82Q.621D}) which  shows 
a large degree of structure, from HI holes and shells (\citealt{1998PASA...15..157M})  
typical of dwarf galaxies (\citealt{1999AJ....118..273W}) to clear indications of a spiral pattern and a central bar. 
The molecular gas traced by the CO(1-0) line shows three well differentiated CO emitting complexes (\citealt{2001AJ....121..727W}) 
related to the main SF complexes. NGC~4214 is a gas rich galaxy: the total mass of atomic gas is 
$M_{\rm HI} = 4.1 \times 10^8$\Msun\ (\citealt{2008AJ....136.2563W}), and the molecular gas mass is 
$M_{\rm H2} = 5.1 \times 10^6$\Msun\ (\citealt{2001AJ....121..727W}, obtained with a Galactic conversion factor).

\begin{figure*}
 \centering
 \includegraphics[width=0.95\textwidth]{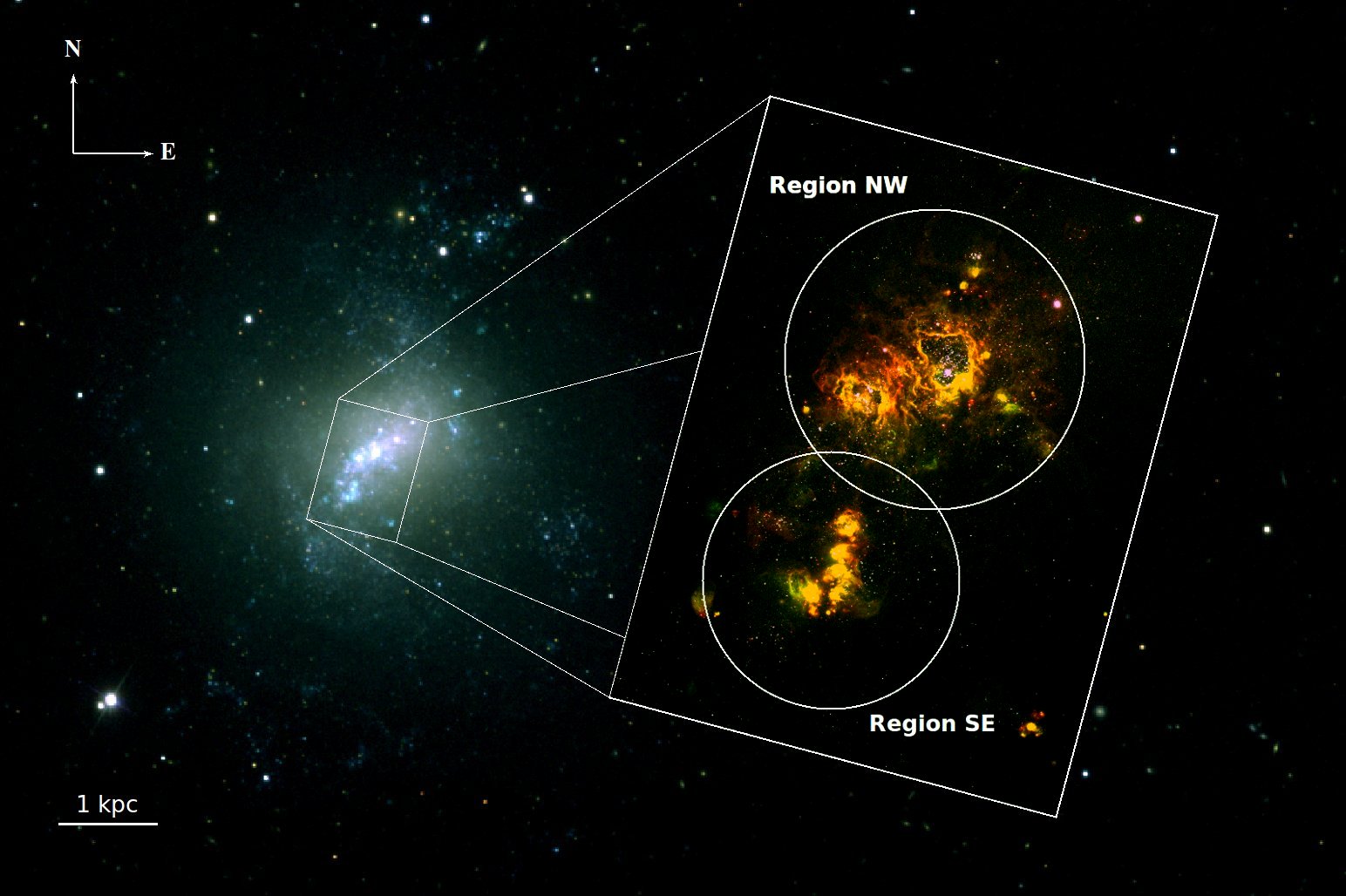}
 \caption{Combined SLOAN and HST image of NGC~4214.
 Large field: Image of NGC~4214 from a RGB combination of SLOAN r, g and u
bands. Zoom-in: Image of the centre of NGC~4214 from a RGB combination of
HST-WFC3 filters F657N ($\rm H_{\alpha} + cont$), F502N ($\rm [OIII] +
cont$) and F336W (cont). The two circles are the apertures that we use for the
aperture photometry of the regions NW and SE. }
  \label{sloan_hst}
\end{figure*}

\citet{2004AJ....127.2031K} reported for NGC~4214 a stellar mass of $\rm \sim 1.5\times10^9$\Msun, similar to the mass found in the Large Magellanic Cloud (LMC). 
Ultraviolet (UV), optical and near-infrared (NIR) images of NGC~4214  show that the young stellar population is embedded in a smooth disk of old stars (see the large field of Fig.\ref{sloan_hst}), which can account for a significant fraction ($\rm\sim75\%$) of the total stellar mass \citep{2011ApJ...735...22W}. Despite this high fraction of old stars, NGC~4214 is a galaxy with an intense recent star formation activity, as shown by the two star forming complexes located in its centre and frequently referred in the literature as NGC~4214-NW and NGC~4214-SE. When resolved, the two complexes show individual smaller knots of star formation (see the small field of Fig.\ref{sloan_hst}). One of the most striking features in these complexes is the large shell structure in the NW region, where most of the gas in front of the central star cluster seems to have been removed by the action of stellar winds and supernovae  (SNe) (\citealt{1998A&A...329..409M}; \citealt{2000AJ....120.3007M}). 
NGC~4214-SE is more compact and shows no clear evidence of a decoupling between the star clusters and the gas. The morphological differences for the two complexes were interpreted as an evolutionary trend by \citet{2000AJ....120.3007M}. Using stellar synthesis models, \citet{2007AJ....133..932U} determined the age, the mass, the radius and the extinction of the star clusters within the NW and SE complexes. They found an age of 5\,Myr for the star clusters in the NW region, whereas the age of the star clusters of the SE region ranges from 1.7 to 4.0\,Myr. 

The internal extinction distribution of the two complexes was studied by \citet{1998A&A...329..409M}. These authors found that the distribution of the dust clouds is correlated with the distribution of the ionised gas: the dust is located at the border of the star clusters in the NW complex, whereas for the SE region the dust clouds seem to be co-spatial with the star clusters. The metallicities of the two complexes have been measured by \cite{1996ApJ...471..211K}, who found values in the range of $Z\sim 0.3\, Z_{\odot}$ with little dispersion, similar to the metallicity of the Small Magellanic Cloud (SMC).

Thus, the large number and the broad range of previous studies performed for this galaxy, as well as the significant amount of multiwavelength data, give us the opportunity of obtain an accurately picture of the stellar, dust and gas components in NGC~4214.

%%%%%%%%%%%%
%%% DATA %%%
%%%%%%%%%%%%

\section{The data}
\label{sec:data}

A wide range of data exists in archives for NGC~4214, from the UV to the submm. The entire dust emission SED from the Mid-IR to the submm is covered by data from SPITZER, HERSCHEL and PLANCK. In addition, to extend the dust SED to the mm range we have mapped the galaxy at 1.2\,mm with the IRAM 30\,m telescope. In the following subsections we describe the data and the reduction that we have performed prior to extracting the photometry. A selection of images at different wavelengths is 
shown in Fig.~\ref{images}.

\subsection{GALEX}

NGC~4214 was observed with \textit{GALEX} the 13th of January 2010 as part of the program GI4-095 (Janice Lee, 2009). Far-ultraviolet (FUV) and near-ultraviolet (NUV) photometric maps were obtained.

The \textit{GALEX} FUV detector has an effective wavelength of 1539$\mbox{\,\AA}$, a bandwidth of 442$\mbox{\,\AA}$ and an image resolution of 4.2\arcsec. For the NUV detector, the effective wavelength is 2316$\mbox{\,\AA}$, the bandwidth  1060$\mbox{\,\AA}$ and the image resolution 5.3\arcsec. With both detectors the two main SF complexes NW and SE are well resolved and the field of view, with a radius of about 36\arcmin, cover the entire disk of NGC~4214. We have retrieved the calibrated maps from the archive and subtracted the background using the corresponding maps from the archive.
  
\subsection{HST-WFC3}

Wide Field Camera 3 (WFC3) data were obtained for NGC~4214 as part of the WFC Science Oversight Committee (SOC) Early Release Science (ERS) program  (program ID11360, P.I. Robert O'Connell). A total of 7 stellar and 7 nebular images were taken using UVIS and IR cameras. Both cameras onboard the HST have a field of view of $\sim$2\arcmin and they cover the central star forming complexes NW and SE, allowing us to separate them into several smaller star forming regions. We obtained calibrated maps   stellar filters F225W, F336W, F438W, F547M, F814W, F110W and F160W
from the HST Data Archive.

\subsection{SPITZER}

The SPITZER data for NGC~4214 used in this work are part of the Local Volume Legacy (LVL) survey (\citealt{2009ApJ...703..517D}). The LVL sample contains 258 galaxies within 11 Mpc, which have been mapped with both MIPS (3 bands) and IRAC (4 bands).

For MIPS data the exposure times were 146.8, 83.8 and 16.76 seconds at 24, 70, and 160\mi, respectively. The area mapped in the three bands, approximately 16\arcmin$\times$16\arcmin, covers the whole disk of NGC~4214. The resolution of 6\arcsec\ for \MIPS{24} allows us to 
determine fluxes individually for the SF complexes NW and SE. This separation is not possible for MIPS 70 and 160\mi\ due to a poorer (18\arcsec\ and 38\arcsec, respectively) angular resolution. We obtained the images from the archive and subtracted the background, which we determined as the mean value of an annulus located outside the disk of the galaxy.

As for several other galaxies in the LVL sample, IRAC data of NGC~4214 was taken from previous SPITZER programs. In the case of NGC~4214, the LVL makes use of the IRAC data from the "Mid-IR Hubble Atlas of Galaxies" (\citealt{2004sptz.prop...69F}). As part of this program, NGC~4214 was observed with IRAC at 3.6, 4.5, 5.8 and 8.0\mi\ in May  2004. In our study of the SF regions, we only use the \IRAC{3.6} band, which we assumed as pure stellar emission, and the \IRAC{8.0} band. In order to isolate the dust component from the \IRAC{8.0} we subtracted the stellar emission  using the \IRAC{3.6} image and applying the
formula provided in  \citet{2004ApJS..154..253H}. With an angular resolution of 1.7\arcsec and 2.0\arcsec\ for the IRAC 3.6 and 8.0\mi\ bands, respectively, the SF complexes NW and SE are resolved into several smaller HII regions.

The total mapped area by IRAC, approximately of 13\arcmin$\times$6\arcmin, covered the central part of the disk of NGC~4214, but not the entire extended disk. Therefore, the total emission at 3.6 and 8\mi\ that we obtain might be underestimated. However, judging from the \IRAC{8.0} image in Fig.~\ref{images}, we believe that the  underestimate is not significant.

The uncertainty in the flux calibration is better than 10\% for all IRAC (\citealt{2005PASP..117..978R}, \citealt{2004ApJS..154...10F}) and MIPS 24\mi\ bands (\citealt{2007PASP..119..994E}), and is better than 20\% for MIPS 70 and 160\mi\ bands (\citealt{2004ApJS..154...25R}). We adopt these numbers as the calibration uncertainties in our error estimates for the fluxes (see Sect.~\ref{sec:ERROR}).

\subsection{HERSCHEL}

NGC~4214 was observed by HERSCHEL's PACS and SPIRE instruments as part of the Dwarf Galaxy Survey programme (PI. S. Madden), a guaranteed time 
key program with the aim of mapping the dust and gas in 51 nearby dwarf galaxies. 

SPIRE data were obtained the 26th of June 2011 and cover a circular area of approximately 17\arcmin. We retrieved the HIPE Level 2.0 images from the archive and performed further analysis for our purposes. Due to their good quality and absence of artefacts, we only subtracted the background which we measured in the same annulus used for MIPS data. The resolution of 18\arcsec\ at \SPIRE{250} 
allowed us to measure individual fluxes for the two star forming complexes. This distinction was not possible in the case of SPIRE 350 and 500\mi\ due to a poorer image resolution (25\arcsec\ and 37\arcsec, respectively). \citet{2010A&A...518L...4S} reported for SPIRE 250, 350 and 500\mi\ an uncertainty in the flux calibration of 15\%, which we adopted in our error calculation.

PACS data were obtained the 27th and 28th of December 2011 and cover an
area of $\sim$25\arcmin$\times$25\arcmin. The data from the archive were processed to level 1 with HIPE 8.2.0. 
To complete the reduction we used
the Scanamorphos \citep{2012arXiv1205.2576R}  map making technique. The
background was subtracted from the final maps using the same annulus
mentioned before. At the resolution of PACS at 70, 100 and 160\mi\
(5.2\arcsec, 7.7\arcsec\ and 12\arcsec, respectively), the two star
forming complexes are resolved. The uncertainty in the flux calibration is
of the order of 10\%.

\subsection{PLANCK}
\label{PlancKData}

The "PLANCK Early Release Compact Source Catalogue" includes three detections of NGC~4214 at 350, 550 and 850\mi, all of them composed of two different observations. For the three detections the subtracted Cosmic Microwave Background (CMB) flux was less than 30\% of the total flux measured originally by PLANCK.
The catalogue gives two estimates for the total flux of a source: i) FLUX, obtained by aperture photometry within the nominal sky-averaged FWHM, which is 4.45\arcmin, 4.71\arcmin\ and 4.62\arcmin\ for 350, 550 and 850\mi, respectively, and ii) FLUXDET, obtained by their native detection algorithm 
\citep{2011A&A...536A...7P}. 
We adopt here the first flux estimate, FLUX, which is recommended for  sources that are point-like with respect to the PLANCK beam, as is the case for
NGC~4214. We use the difference between the two estimates (which is larger than the nominal error of each flux) as an estimate of the errors.

\subsection{IRAM 30\,m}

Several observations at 1200\mi\ were made by our group between December 2009 and November 2010 at the IRAM 30\,m telescope on Pico Veleta (Spain), with the 37-channels bolometer array of the Max-Planck-Institut f\"ur Radioastronomie (MPIfR). The 37 pixels are located in a hexagonal structure with a beam-size of 10.8\arcsec\ and a pixel-to-pixel separation of about 23\arcsec. The observations were done on-the-fly and they were calibrated by observations of the planet Mars and secondary calibrators. Observations were reduced and combined into an image with Robert Zylka's MOPSIC\footnote{See http://www.iram.es/IRAMES/mainWiki/CookbookMopsic} pipeline in a standard manner including baseline subtraction, spike and sky noise removal. We smoothed the combined image to a resolution of 18\arcsec\ in order to increase the signal-to-noise ratio. At this resolution, the two main star forming complexes are still resolved. 
Our observations were not sensitive enough to detect any extended emission outside the two main SF complexes. We adopt 30\% calibration uncertainty.

\subsection{Thermal radio emission}
\label{sec:VLA}

The thermal radio emission in an HII region  is proportional to the production rate of Lyman continuum photons $N_{\rm Lyc}$. \citet{1992ARA&A..30..575C} gives the following  relation between the thermal radio emission and
the extinction-corrected \hbeta\ line flux,  $F$(\hbeta), which is proportional to  $N_{\rm Lyc}$:

\begin{equation}
\label{eq:CONDON}
 \frac {S_{\rm T}}{\rm mJy} \sim {\frac{1}{0.28}}{\left(\frac{T}{10^4 {\rm K}}\right)}^{0.52}{\left(\frac{\nu}{\rm GHz}\right)}^{-0.1}{\frac{F({\rm H\beta})}{\rm 10^{-12} erg\, cm^{-2} s^{-1}}},
\end{equation}
where $S_{T}$ is the thermal flux density at the frequency $\nu$ for an HII region with a temperature $T$.

We use the measurements of the \halpha\ emission from \citet{2000AJ....120.3007M} to calculate the expected thermal radio emission from NW and SE.
They  presented an \halpha\ narrowband map of the centre of NGC~4214 obtained with the HST Wide Field and Planetary Camera (WFPC2).
They used the spectroscopic values of the Balmer ratio \halpha/\hbeta\ presented by \citet{2000PASP..112.1138M} to correct for dust extinction and they reported intrinsic \halpha\ fluxes of NW and SE. Assuming that the dust is well mixed with the  gas, \citet{2000AJ....120.3007M} calculated extinction-corrected H$\alpha$ fluxes of $12.88\times10^{-12}$ erg cm$^{-2}$ s$^{-1}$ for the NW complex and $5.75\times10^{-12} $erg cm$^{-2}$ s$^{-1}$ for the SE complex. 

For the NW region, \citet{2000AJ....120.3007M} used a circular aperture of radius 20.6\arcsec, practically identical to the aperture used in this work (see Sec.~\ref{sec:APERT}). For the SE region they used a smaller aperture (9.96\arcsec). We therefore multiplied their flux by a factor 1.4, corresponding to the ratio of the uncorrected \halpha\ fluxes measured from the same image in apertures of 9.96\arcsec and 18\arcsec, respectively. Using these extinction-corrected \halpha\  fluxes, the intrinsic $H\alpha/H\beta$ ratio and the temperature of $10^4$K reported by \citet{1996ApJ...471..211K}, we obtain with Eq.~\ref{eq:CONDON} flux densities at 8.46 GHz (3.5 cm) of 13 and 8 mJy for NW and SE, respectively.

The combined prediction for the thermal radio emission of NW and SE (21 mJy) can be roughly compared with the measurement from the VLA at 8.46 GHz of 24.2$\pm4.8$~mJy given by \citet{2011ApJ...736..139K} for the total radio emission, which includes synchrotron and thermal emission. From the uncorrected \halpha\ emission, \citet{2011ApJ...736..139K} estimated that the fraction of thermal emission is on average $\sim 0.5$ (see Fig.~9 in their paper). However, since these authors did not correct the \halpha\ emission for dust attenuation, their estimation gives a lower limit of the thermal fraction. Our estimation of free-free emission (21~mJy) falls within the limits of thermal fractions of 0.5 (12.1~mJy) and 1.0 (24.2~mJy) and is thus consistent with their data.
We used their limits to roughly estimate an error of the thermal radio flux of $\rm\sim30\%$.  

%\textbf{Although these values were obtained using extinction corrections independent of the radiation transfer calculations of this paper, this inconsistency is inconsequential for our radiation transfer analysis due to the relatively small amplitude of the radio free-free fluxes compared to the dust emission.}

\begin{figure*}
 \centering
 \vspace{0.3cm}
 \includegraphics[width=0.95\textwidth]{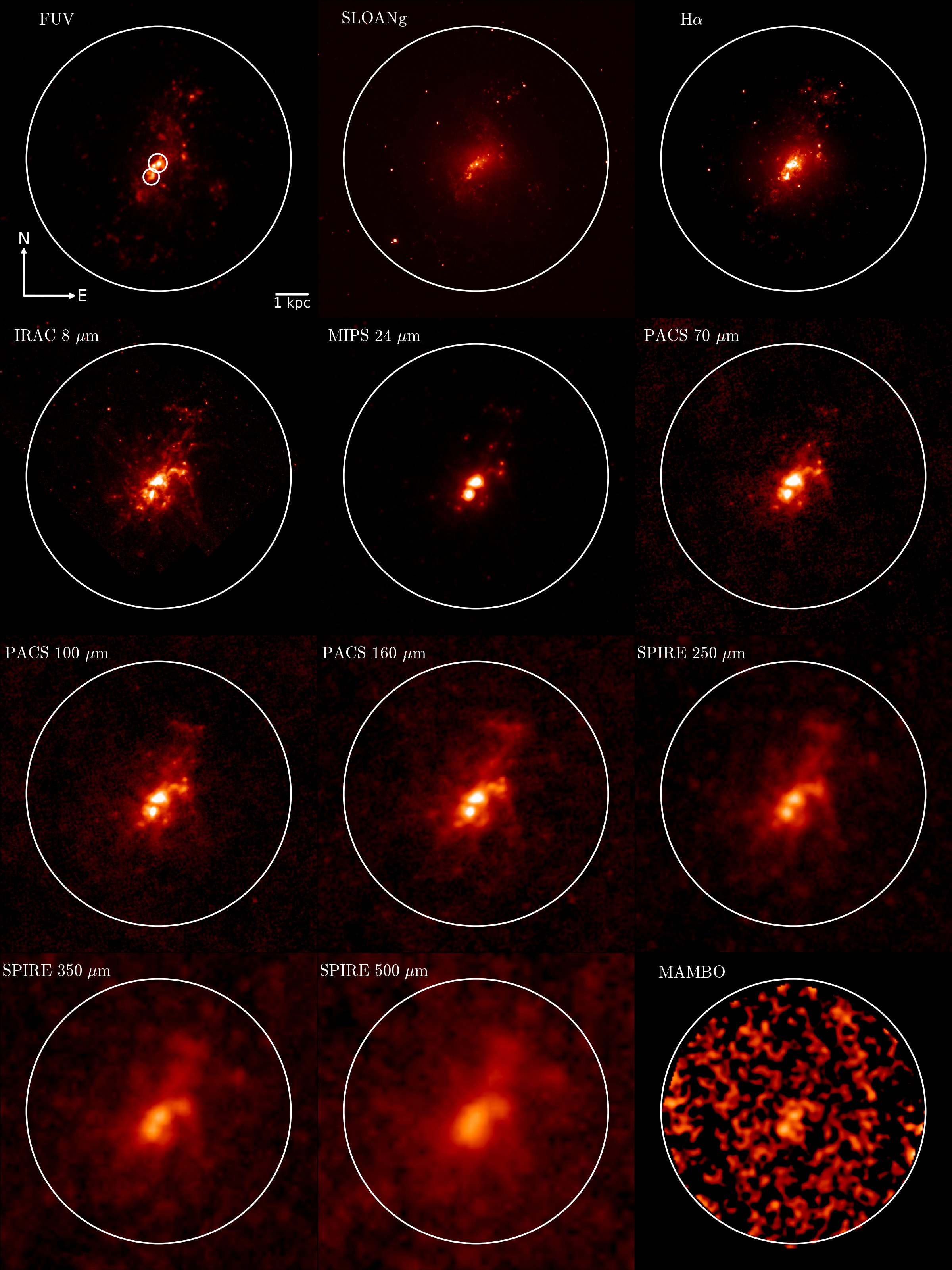}
 \caption{Images of NGC~4214 at different wavelengths. The large circle shows the aperture that we chose for the entire disk emission and 
 the smaller circles in the top-left panel the aperture to measure the emission from the HII regions NW and SE. }
  \label{images}
\end{figure*}

%%%%%%%%%%%%%%%%%%%%
%%%% PHOTOMETRY %%%%
%%%%%%%%%%%%%%%%%%%%

\section{Photometry}
\label{photometry}

We need to separate the dust emission from the two major SF complexes NW and SE and the diffuse disk, in order to establish the individual SEDs of these morphological components.
We therefore independently carried out aperture photometry for the regions SE and NW and for the total emission of the entire galaxy.
We determined the diffuse dust emission as the difference between the total emission and the sum of the emissions of SE and NW regions. 

Prior to extracting the photometry, we regridded all the images to a common pixel size keeping the original resolution of each image. Then, corresponding aperture corrections were applied to the fluxes in each band. 
In this section we describe how we performed the aperture photometry, which corrections were applied and how the errors were handled.

\begin{table*}
\vspace{0.40cm}
   
%Table generated with dwarfSED.pro on Mon Jul 30 19:11:56 2012

%\begin{table*}
\centering
\begin{tabular}{c c c c c c c} \hline \noalign{\medskip}
\textbf{BAND} & $\lambda_{0}$($\mu$m) & $F_{\rm NW}^{\rm sub}$ (Jy) & $C_{\rm apert}$ & $C_{\rm color}$ & $F_{\rm NW}$ (Jy) \\ \noalign{\smallskip} \hline \noalign{\medskip}
%%%%%%%%%%%%%%%%%%%%%%%%%%%%%%%%%%%%%%%%%%%%%%%%%%%%%%%%%%%%%%%%%%%%%%%%%%%%%%%%%%%
IRAC8 & 7.872 & 0.163 & 1.000 & 0.590 & 0.276 $\pm$ 0.035 \vspace{0.15cm} \\ 
MIPS24 & 23.680 & 0.733 & 1.078 & 0.986 & 0.801 $\pm$ 0.103 \vspace{0.15cm} \\ 
PACS70 & 70.000 & 7.612 & 1.039 & 0.982 & 8.053 $\pm$ 1.031 \vspace{0.15cm} \\ 
PACS100 & 100.000 & 8.443 & 1.050 & 0.985 & 9.000 $\pm$ 1.153
 \vspace{0.15cm} \\ 
PACS160 & 160.000 & 6.114 & 1.075 & 1.010 & 6.507 $\pm$ 0.833
 \vspace{0.15cm} \\ 
SPIRE250 & 250.000 & 2.329 & 1.110 & 0.992 & 2.605 $\pm$ 0.443
 \vspace{0.15cm} \\ 
MAMBO$^{(*)}$ & 1200.000 & 0.031 & 1.067 & 1.000 & 0.033 $\pm$ 0.010 \vspace{0.15cm} \\ 
\condon & 5000.000 & 0.011 & 1.000 & 1.000 & 0.011 $\pm$ 0.003
 \vspace{0.15cm} \\ 
%%%%%%%%%%%%%%%%%%%%%%%%%%%%%%%%%%%%%%%%%%%%%%%%%%%%%%%%%%%%%%%%%%%%%%%%%%%%%%%%%%%
\hline \\
\end{tabular}
%\label{DustNW}
%\caption{Region NW photometric points.}
%\end{table*}

   \caption{Flux densities for region NW obtained for an aperture of 21\arcsec\ radius centred at RA$=$\RA{12}{15}{39.6} Dec$=$\DEC{+36}{19}{36}{5} (J2000) (see Figs~\ref{sloan_hst} and \ref{images}). $\,F_{\rm NW}^{\rm sub}$ is the measured, background-subtracted flux and $F_{\rm NW}$ is the final flux, after applying aperture and colour corrections. $^{(*)}$ The contribution from thermal radio emission was not subtracted.\label{tab:DustNW}}  
\vspace{0.40cm}
   
%Table generated with dwarfSED.pro on Mon Jul 30 19:11:56 2012

%\begin{table*}
\centering
\begin{tabular}{c c c c c c c} \hline \noalign{\medskip}
\textbf{BAND} & $\lambda_{0}$($\mu$m) & $F_{\rm SE}^{\rm sub}$ (Jy) & $C_{\rm apert}$ & $C_{\rm color}$ & $F_{\rm SE}$ (Jy) \\ \noalign{\smallskip} \hline \noalign{\medskip}
%%%%%%%%%%%%%%%%%%%%%%%%%%%%%%%%%%%%%%%%%%%%%%%%%%%%%%%%%%%%%%%%%%%%%%%%%%%%%%%%%%%
IRAC8 & 7.872 & 0.105 & 1.000 & 0.590 & 0.178 $\pm$ 0.023 \vspace{0.15cm} \\ 
MIPS24 & 23.680 & 0.602 & 1.069 & 0.986 & 0.653 $\pm$ 0.084 \vspace{0.15cm} \\ 
PACS70 & 70.000 & 4.659 & 1.032 & 0.982 & 4.896 $\pm$ 0.627 \vspace{0.15cm} \\ 
PACS100 & 100.000 & 5.416 & 1.040 & 0.985 & 5.719 $\pm$ 0.732
 \vspace{0.15cm} \\ 
PACS160 & 160.000 & 4.094 & 1.057 & 1.010 & 4.285 $\pm$ 0.549
 \vspace{0.15cm} \\ 
SPIRE250 & 250.000 & 1.689 & 1.078 & 0.992 & 1.835 $\pm$ 0.312
 \vspace{0.15cm} \\ 
MAMBO$^{(*)}$ & 1200.000 & 0.029 & 1.021 & 1.000 & 0.029 $\pm$ 0.009 \vspace{0.15cm} \\ 
\condon & 5000.000 & 0.005 & 1.400 & 1.000 & 0.007 $\pm$ 0.002
 \vspace{0.15cm} \\ 
%%%%%%%%%%%%%%%%%%%%%%%%%%%%%%%%%%%%%%%%%%%%%%%%%%%%%%%%%%%%%%%%%%%%%%%%%%%%%%%%%%%
\hline \\
\end{tabular}
%\label{DustSE}
%\caption{Region SE photometric points.}
%\end{table*}

   \caption{Flux densities for region SE obtained for an aperture of 18\arcsec\ radius centred at RA$=$\RA{12}{15}{40.8} Dec$=$\DEC{+36}{19}{05}{5} (J2000) (see Figs~\ref{sloan_hst} and \ref{images}). $\,F_{\rm SE}^{\rm sub}$ is the measured, background-subtracted flux and $F_{\rm SE}$ is the final flux, after applying aperture and colour corrections. $^{(*)}$ The contribution from thermal radio emission was not subtracted.\label{tab:DustSE}}
\vspace{0.40cm}
% \end{table*}
% \begin{table*}
   
%Table generated with dwarfSED.pro on Wed Aug 22 19:38:24 2012

%\begin{table*}
\centering
\begin{tabular}{c c c c c c c} \hline \noalign{\medskip}
\textbf{BAND} & $\lambda_{0}$($\mu$m) & $F_{\rm DISK}^{\rm sub}$ (Jy) & $C_{\rm color}$ & $F_{\rm DISK}$ (Jy) & $F_{\rm DIFF}$ (Jy) \\ \noalign{\smallskip} \hline \noalign{\medskip}
%%%%%%%%%%%%%%%%%%%%%%%%%%%%%%%%%%%%%%%%%%%%%%%%%%%%%%%%%%%%%%%%%%%%%%%%%%%%%%%%%%%
IRAC8 & 7.872 & 0.678 & 0.590 & 1.149 $\pm$ 0.115 & 0.695$^{+0.075}_{-0.090}$
 \vspace{0.15cm} \\ 
MIPS24 & 23.680 & 2.050 & 0.986 & 2.080 $\pm$ 0.208 & 0.625$^{+0.104}_{-0.124}$
 \vspace{0.15cm} \\ 
MIPS70 & 71.440 & 23.945 & 0.893 & 26.814 $\pm$ 5.363 & 
13.417$^{+3.082}_{-4.084}$ \vspace{0.15cm} \\  
PACS70 & 70.000 & 24.209 & 0.982 & 24.652 $\pm$ 2.466 & 
11.703$^{+1.393}_{-1.672}$ \vspace{0.15cm} \\ 
PACS100 & 100.000 & 35.309 & 0.985 & 35.847 $\pm$ 3.585 & 
21.128$^{+2.279}_{-2.735}$ \vspace{0.15cm} \\ 
PACS160 & 160.000 & 34.097 & 1.010 & 33.759 $\pm$ 3.376 & 
22.967$^{+2.380}_{-2.856}$ \vspace{0.15cm} \\ 
MIPS160 & 155.900 & 39.016 & 0.971 & 38.574 $\pm$ 7.715 & 
27.287$^{+5.605}_{-6.042}$ \vspace{0.15cm} \\ 
SPIRE250 & 250.000 & 18.888 & 0.992 & 19.033 $\pm$ 2.855 & 
14.594$^{+2.204}_{-2.645}$ \vspace{0.15cm} \\ 
PLANCK350 & 350.000 & 9.546 & 0.986 & 9.682 $\pm$ 0.988 & 
7.933$^{+0.797}_{-0.870}$ \vspace{0.15cm} \\ 
SPIRE350 & 363.000 & 10.059 & 0.999 & 10.068 $\pm$ 1.510 & 
8.453$^{+1.281}_{-1.321}$ \vspace{0.15cm} \\ 
SPIRE500 & 517.000 & 4.497 & 1.025 & 4.387 $\pm$ 0.659 & 
3.848$^{+0.580}_{-0.590}$ \vspace{0.15cm} \\ 
PLANCK550 & 550.000 & 3.635 & 0.921 & 3.947 $\pm$ 1.318 & 
3.510$^{+1.173}_{-1.177}$ \vspace{0.15cm} \\ 
PLANCK850 & 850.000 & 0.946 & 0.887 & 1.066 $\pm$ 0.264 & 
0.958$^{+0.231}_{-0.232}$ \vspace{0.15cm} \\ 
MAMBO$^{(*)}$ & 1200.000 & 0.350 & 1.000 & 0.350 $\pm$ 0.106 & 0.288$^{+0.087}_{-0.105}$
 \vspace{0.15cm} \\ 
%%%%%%%%%%%%%%%%%%%%%%%%%%%%%%%%%%%%%%%%%%%%%%%%%%%%%%%%%%%%%%%%%%%%%%%%%%%%%%%%%%%
\hline \\
\end{tabular}
%\label{DustDISK}
%\caption{Region DISK and diffuse photometric points.}
%\end{table*}

   \caption{Flux densities of NGC~4214 for an aperture of 5\arcmin\ radius centred at  RA$=$\RA{12}{15}{39.1} Dec$=$\DEC{+36}{19}{34}{5} (J2000) (see Fig.~\ref{images}). $\,F_{\rm DISK}^{\rm sub}$ is the measured, background-subtracted flux, $F_{\rm DISK}$ is the final flux of the entire galaxy, after applying colour correction, and $F_{\rm DIFF}$ is the flux of the diffuse emission obtained as $F_{\rm DISK} - (F_{\rm NW} + F_{\rm SE})$. The lower error of $F_{\rm DIFF}$ includes the contribution of smaller and less intense SF regions in the disk (see Sect.~\ref{sec:RESULTS}). $^{(*)}$ The contribution from thermal radio emission was not subtracted.\label{tab:fluxes_diffuse}}
\vspace{0.45cm}
\end{table*}

\subsection{Apertures}
\label{sec:APERT}

We chose the apertures around the HII regions large enough to enclose the entire localized dust emission
and simultaneously tried to include as little as possible emission from the diffuse component.
Mainly based on the MIPS 24\mi\ image, we chose apertures with radii of 21\arcsec\ and 18\arcsec\ for the NW and SE complexes (see Figures~\ref{sloan_hst} and \ref{images}), respectively. 
The 24~\mi\  image was used as it is the filter which is expected to most faithfully trace the warm dust in thermal equilibrium with the strong radiation fields from the star clusters in the PDRs and HII regions associated with the SF regions.
We were able to measure the emission from each SF complex only for those wavelengths where both complexes were spatially resolved, which excluded the maps of MIPS 70 and 160~\mi\ and SPIRE 350 and 500\mi.
The measured, background-subtracted fluxes are listed in column 3 in Tables~\ref{tab:DustNW} and \ref{tab:DustSE} for NW ($F_{\rm NW}^{\rm sub}$) and SE ($F_{\rm SE}^{\rm sub}$), respectively. 

One issue of particular relevance to the quantitative analysis presented in this paper is that of the subtraction of the local background underlying the NW and SE SF regions. Ideally, we need  to cleanly separate the localised dust/PAH emission associated  with the HII regions and the local PDRs on scales of tens of parsecs (marking the interface between the tenuous gas ionised by the star  cluster and the dense material of the parent molecular clouds) from the underlying extended emission from diffuse dust distributed on the kpc scale of the disk. 
%This is because these are the two types of structures for which our radiation transfer model predicts an SED. 
In practice it is very challenging to perform such a separation, not only because of blending effects due to the finite angular resolution (these are quantified through the aperture corrections described in Sect. 3.2, below) but also because of the limited range of UV photons which have escaped the SF regions and have passed into the diffuse dust layer, forming a halo around the SF regions.
%In the case of the very luminous and  dominant NW and SE SF regions, this escaping UV light is expected to produce a detectable halo of dust emission extending beyond the local PDRs on linear scales of the mean free path between absorption events on dust grains in the diffuse dust layer.}
%
%\textbf{Direct evidence for this phenomenon is most particularly shown by the relatively high resolution 8 micron IRAC images shown in Fig. 2, which clearly show evidence for PAH emission associated with the NW and SE SF regions, but extending well beyond the SF regions  delineated by \halpha emission on the HST images as shown in Fig. 1. Although prominence of the halo in relation to the 8 emission from the star formation region at 8 micron may in part be due to the suppression of PAH emission from the SF region itself (this is further explored in Sect. 6.1), it also consitututes evidence for a strong illumination of the diffuse dust layer by non-ionising UV escaping the SF region. As such, one would also this same UV radiation field to give rise to haloes at other wavelengths, most especially in the FIR.}
%

%This halo  from dust/PAH should in principle be attributed to the diffuse dust emission component of the galaxy when accounting for it in the radiation transfer modelling of the galaxy. In practice, however, one expects 
The extent and brightness of this diffuse halo emission is expected to vary strongly with infrared wavelength in a way which is both difficult to measure directly %(due  to the marginal angular resolution at the longer wavelengths), 
and also uncertain to predict theoretically on an a priori basis.
% (since the radial scale and brightness of the dust emission in the halo will depend on the local volume density of grains in the diffuse disk, which in turn is  determined by the face-on opacity of the disk and the adopted ratio of scalelength- to-scaleheight of the diffuse dust). 
Therefore, for simplicity, we have opted in this paper to make no attempt to subtract the underlying diffuse background when performing aperture photometry on the SF regions. For future reference we simply note here that we may thereby be overestimating the flux densities of the NW and SE SF regions, most particularly in the PAH bands, but also to some extent around the peak of the SED in the FIR. The effect at 24~\mi , however, is likely to be small, since, as already noted, this band is primarily sensitive to warm dust heated by the intense radiation fields in the PDRs and HII regions close to the star clusters.

For the measurement of the total emission of NGC~4214 we chose an aperture of 5\arcmin\ in radius (see Figure~\ref{images}). We tested with growth curves at all wavelengths that this aperture enclosed the entire emission of the galaxy.
For the PLANCK measurements, which include the total emission of the galaxy, the apertures could not be chosen. However, their values are very close (see Section \ref{PlancKData}) to ours, and our growth curves showed that at the radius  of the PLANCK beam  the emission was enclosed completely.

The total emission from the MAMBO 1.2\,mm map is more uncertain due to the poor signal-to-noise ratio at the outer parts of the area covered by our observations, which is practically  the same as the aperture of 5\arcmin\ used for the other images. In the moderately smoothed image (resolution 18\arcsec , shown in Fig. 2) no 
diffuse emission can be seen. However, when 
 we smooth the map to an angular resolution of 40\arcsec\  diffuse dust emission becomes visible  in the inner 160\arcsec\ whose
 structure corresponds very well to the  SPIRE 500\,\mi\ map. Beyond this radius we are not sure that the structures are real. 
 The growth curve of MAMBO keeps raising until $\rm\sim200$\arcsec\ where it becomes approximately flat. 
 We obtained integrated fluxes of  $\rm0.26\pm0.03$ Jy and $\rm0.35\pm0.11$~Jy for the apertures of 160\arcsec\ and 300\arcsec, respectively. 
The lower value should represent a reliable lower limit for the total flux in NGC~4214. The total flux is more uncertain, however,
given the  flatness of the growth-curve
 beyond 200\arcsec, we are therefore confident that the flux at 300\arcsec\ is a reasonable estimate. 
Due to these uncertainties,  
we did not include the MAMBO data point in our fitting procedure but we show it 
in the figures for comparison with the models.
 
The measured, background-subtracted  fluxes of the  total emission are listed in column 3 of Table~\ref{tab:fluxes_diffuse}.

\subsection{Aperture correction}
\label{sec:AC}

Because of the different resolutions of the images we applied aperture corrections to the NW and SE IR fluxes.
%  in order to obtain the total flux enclosed in the selected apertures in each band. 
%We applied aperture corrections to the fluxes derived with aperture photometry for the  complexes NW and SE from the images of MIPS, SPIRE and MAMBO, where the resolution was a considerable fraction of  the aperture size.
We derived values of the aperture correction by adopting the \IRAC{8} as the high-resolution model of the flux distribution.
We first measured the flux within the NW and SE apertures in the \IRAC{8} image.
Then we convolved the \IRAC{8} image with the point-spread-function (PSF) of MIPS~24, PACS~70, PACS~100, PACS~160, SPIRE~250 and MAMBO, and measured again the flux within the same apertures.
The ratio between the flux measured in the original and convolved images  gives the aperture correction for each band. 
The values are listed in column 4 of Tables~\ref{tab:DustNW} and \ref{tab:DustSE}.

\subsection{Colour corrections}
\label{sec:CC}

In order  to directly compare  the monochromatic fluxes with the models of \citet{2011A&A...527A.109P} and \citet{2008ApJS..176..438G}, we applied colour corrections to the IRAC, MIPS, PACS, SPIRE and PLANCK fluxes. We followed the procedures described in the corresponding manuals\footnote{see the \href{http://irsa.ipac.caltech.edu/data/SPITZER/docs/irac/iracinstrumenthandbook/home/}{IRAC Instrument Handbook}, \href{http://irsa.ipac.caltech.edu/data/SPITZER/docs/mips/mipsinstrumenthandbook/51/}{MIPS Instrument Handbook},
\href{http://herschel.esac.esa.int/Docs/PACS/html/pacs_om.html}{PACS Observer's Manual}, 
\href{http://herschel.esac.esa.int/Docs/SPIRE/html/spire_om.html}{SPIRE Observers' Manual} and 
\citet{2011yCat.8088....0P} for IRAC, MIPS, SPIRE and PLANCK, respectively.}, where colour correction factors for different source distributions are listed.

We chose the colour correction factors for the spectrum that  most closely resembled the model spectrum at a given wavelength.
For the MIPS 24\mi\ band we adopted a blackbody with a 
temperature of 70\,K, for the MIPS 70\mi, MIPS 160\mi\ and PACS bands a blackbody with a 
temperature of 50\,K, for SPIRE bands and PLANCK 350\,\mi\ we chose  
a power-law spectrum with  $\alpha=+2.0$, for PLANCK 550\,\mi\ a power-law spectrum with  $\alpha=+2.5$, and for PLANCK 850\,\mi\ a power-law spectrum with $\alpha=+3.0$ (where the exponent $\alpha$ is defined as $F_\nu \propto \nu^{\alpha}$). 
For the particular case of IRAC 8~\mi , we adopted the colour correction for a PAH-dominated spectrum. 
At 8~\mi\  we additionally compared model and data by generating synthetic integrated fluxes. For this we (i) multiplied the model spectrum 
with the 8~\mi\ filter profile  and integrated it over the 8~\mi\  filter band and (ii)   multiplied the observed 8~\mi\ data, adopting the black body of $\rm T=10^4\,K$ profile
on which the flux definition of IRAC is based, with  the 8~\mi\ profile and integrated it  over the  8~\mi\  filter band. The comparison of both values allowed an alternative 
estimate of the goodness of the fit, based entirely on the data and the model spectrum.
We found that this comparison  produced within 10\% the same  ratios as the comparison between
colour corrected data point and model flux.

The colour corrections are listed in Tables~\ref{tab:DustNW}, \ref{tab:DustSE} and \ref{tab:fluxes_diffuse}.

\subsection{Line contamination}
\label{sec:LC}

Most of the FIR/submm broad-band filters used for this work are contaminated by molecular or atomic emission lines. In the cases where data were available, we estimated the line contamination by comparing the luminosity of the emission line with the total luminosity measured in the filter, obtained by integrating 
 the product of the source spectrum and the spectral response of the filter over the frequency range of the filter band.

Emission lines from different CO rotational transitions fall into SPIRE, PLANCK and MAMBO filters. 
We used the %Owens Valley Radio Observatory (OVRO) 
\nobreak{CO(1-0)}
 map presented by \citet{2001AJ....121..727W} as a reference. We determined a velocity integrated flux of 9.3  Jy \kms\ in the NW region and 12.8 Jy \kms\ in the
SE region. The MAMBO filter band, with a bandwidth of about 80 GHz, is affected by the 
CO(2-1) line. 
We estimate the flux of the CO(2-1) line from  the measured CO(1-0) value  assuming that the line intensity ratio it is given by the value for 
optically thick, thermalised CO,
$I_{\rm CO(1-0)}/I_{\rm CO(2-1)} = 1$. 
%Since there is no available data for these CO transitions, we assumed as an upper limit  that they have the same luminosity as CO(1-0).
The contribution of the CO(2-1) line to the total measured flux at 1.2\,mm  of the NW and SE complex
is then  0.87\% and 1.27\%, respectively. For other bands, the contamination by higher CO transitions is more than an one order of magnitude smaller.

PACS spectroscopic maps of atomic FIR lines ([OI] 63, [OIII] 88, [NII] 122,  [OI] 146, [CII] 158 and [NII] 205\mi) of NGC~4214 were presented by \citet{2010A&A...518L..57C}. With a field of view of 1.6\arcmin\ x 1.6\arcmin, only the central part of the galaxy is observed. 
However, this region completely covers the main SF complexes NW and SE where the major part of the line emission  originates. Therefore, we do not expect to severely 
underestimate the total emission of these lines.
[OI] 63 and [OIII] 88\mi\ fall into the bandpass of the MIPS~70 and PACS~70 filter. The luminosities reported by \citet{2010A&A...518L..57C} are $0.89\times10^{6}~L_{\hbox{$\odot$ }}$ for [OI] 63 $\mu $m and $1.99\times10^{6}~L_{\hbox{$\odot$ }}$ for [OIII] 88\mi. The sum of these values corresponds to
 1.11\% (1.33\%) of the total luminosity measured in the PACS~70 (MIPS~70) band.  
[OIII] 88\mi\ fall into the bandpass of the PACS~100 filter. Its luminosity corresponds to 1.73\% of the total luminosity measured in the PACS~100 band.  
[OI] 146 and [CII] 158\mi\ fall into the bandpass of the MIPS~160 and PACS~160 filter. The reported luminosities for [OI] 146 and [CII]\mi\ lines are $0.05\times10^{6}~L_{\hbox{$\odot$ }}$ and $2\times10^{6}~L_{\hbox{$\odot$ }}$, respectively, which together represents  2.88\% (4.26\%) of the total luminosity 
measured in the PACS~160 (MIPS~160) band.

Due to the small contribution from the atomic and molecular lines to the FIR/submm bands, we did not apply any decontamination to our measurements except for the PACS~160 and MIPS~160 bands. These have been included in Tables~\ref{tab:DustNW}, \ref{tab:DustSE} and \ref{tab:fluxes_diffuse}.

\subsection{Error handling}
\label{sec:ERROR}

%In our analysis, the monochromatic flux for each band, $F_{reg}$, with the subindex \textit{reg} corresponding to \textit{NW}, \textit{SE} or \textit{disk} depending on the aperture considered, was calculated from the background-subtracted flux as it follows:

%\begin{equation}
% F_{reg} = \frac{C_{\rm apert}}{C_{colour}} \times F_{reg}^{\rm sub} - F^{lines},
%\end{equation}
%
%where $C_{\rm apert}$ and $C_{colour}$ are the aperture and colour corrections, respectively, and $F^{lines}$ is the total flux of the lines present in the bandpass of the filter. %To estimate the flux of the background, we measure the "average value" in a ring beyond the disk of the galaxy

In our error analysis we took into account three types of error: (i) calibration, $\Delta_{\rm cal}$, for which we adopted the values mentioned in Sect.~\ref{sec:data}, (ii) measurement error due to background fluctuations, $\Delta_{\rm back}$, and (iii) an estimate of the error due to
the uncertainty in the aperture size of regions SE and NW, $\Delta_{\rm apert}$ (this error was not relevant for the total emission). We neglected the uncertainties introduced by the colour and aperture corrections.

The error due to the background fluctuations was calculated assuming that each pixel  within the aperture has an error given by
the standard deviation of the background noise, $\sigma_{\rm back}$. 
In addition, we had to take into account the error of the background which was 
subtracted within an
aperture of $N_{\rm apert}$ pixels. This error is  $\sigma_{\rm back} N_{\rm apert} /\sqrt{N_{\rm back}}$, 
where $N_{\rm back}$ is the number of pixels used to compute the level of background.
This gives (see also  \citealt{2012ApJ...745...95D}) a total error for the background subtracted flux of: 
%(see also  \citealt{2012ApJ...745...95D}, but note that we use units of Jy/pixel):

\begin{equation}
  \Delta_{\rm back} =  \sigma_{\rm back}  \sqrt{N_{\rm apert} + \frac{N_{\rm apert}^2}{N_{\rm back}} }
\end{equation}

We estimated the error $\Delta_{\rm apert}$ by changing the aperture size for the regions NW and SE by $\pm$1\arcsec. We obtained
changes in the integrated fluxes between 4 and 8\% for the different bands. We adopted conservatively an
error of  $\Delta_{\rm apert}=8\%$ for all bands. 

The final error for the flux is the quadratic sum of $\Delta_{\rm cal}$, $\Delta_{\rm back}$, and $\Delta_{\rm aper}$. This is listed in Tables~\ref{tab:DustNW}, \ref{tab:DustSE} and \ref{tab:fluxes_diffuse} for the fluxes in each band.
%
%\begin{equation}
% \Delta F_{reg}  = \frac{ C_{\rm apert} }{ C_{colour} } \times \sqrt{ \Delta_{\rm cal}^2 + \Delta_{\rm back}^2 + \Delta_{\rm apert}^2}
%\end{equation}
%In the small apertures used to calculate the emission of the HII regions, 
The dominant error sources  were $\Delta_{\rm cal}$ and $\Delta_{\rm aper}$, whereas $\Delta_{\rm back}$ was found to be negligible for most wavelengths.

%%%%%%%%%%%%%%%%
%%%% MODELS %%%%
%%%%%%%%%%%%%%%%

\section{Models for the dust emission}

We analyse the full UV to FIR/submm SED of the different emission components of the galaxy using the radiation transfer model of \citet{2011A&A...527A.109P}, which self-consistently treats the dust emission from diffuse and SF complexes components, considering the illumination of diffuse dust both by the distributed stellar populations, and by escaping light from the HII regions. While maintaining consistency with the framework of \citet{2011A&A...527A.109P} model, we use the model of \citet{2008ApJS..176..438G} to provide a detailed description of the dust emission from the two central SF complexes NW and SE. In the following sections we present a brief description of the physics and parameters of these models.

\subsection{The model of Popescu et al. (2011)}
\label{DIFFmodels}

\citet{2011A&A...527A.109P} presented a self-consistent model based on full radiative transfer calculations of the propagation of starlight in disk galaxies. To approximate the large scale geometry of the galaxy \citep[see Fig.~1 in][]{2011A&A...527A.109P}, they used two separate components: (i) an old component consisting of an old stellar disk, an old stellar bulge and a dust disk, and (ii) a young component consisting of  a young stellar disk and a dust disk. The young component is introduced to mimic the more complex distribution of young stars and diffuse dust associated with the spiral arms. Apart from the diffuse component, the model includes a clumpy component, consisting of the parent molecular clouds of massive stars. The input parameters of this model are:

\begin{itemize}
 \item The total central face-on B-band opacity, \taubf , which is the sum of the central face-on B-band  opacities of the young and old dust disks.
 \item The star formation rate, $SFR$.%, which is derived from the deattenuated UV-to-optical luminosity.
 \item The clumpiness factor $F$, which can be physically identified with the luminosity-weighted mean fraction of directions from the massive stars, averaged over the lifetime of the stars, which intersects the birth-cloud.  The clumpiness factor $F$ is linked to the  fraction of photons that escapes (\fesc) from the SF regions into
 the diffuse medium as $F=1-$\fesc.
 \item The normalized luminosity of the old stellar disk, $old$.%, which is derived from deattenuated optical-to-near-infrared luminosity.
 \item The bulge-to-disk ratio, $B/D$, which determines the bulge contribution to the old stellar radiation field.
 \item The radial scale-length of the old stellar disk, $h_s$, which defines the size of the galaxy.
 All other spatial scales in the galaxy (scale-length of the young stellar population, scale-length of the dust and 
 vertical scale-heights for the different components) have a constant ratio with $h_s$ \citep[see Table E.1 in][]{2011A&A...527A.109P}.
 \item The inclination angle, $i$.

\end{itemize}

From the primary parameters $SFR$ and $F$, \citet{2011A&A...527A.109P} define the star formation rate powering the diffuse emission, $SFR'$, as it follows (Eq.~45 in \citealt{2011A&A...527A.109P}):

\begin{equation}
 SFR'= SFR \times \left( 1-F \right)
\end{equation}

The library of diffuse SEDs of \citet{2011A&A...527A.109P} contains results for a four-dimensional parameter space  spanned by
\taubf , $SFR'$, $old$ and $B/D$. The diffuse component is calculated as an extrinsic quantity corresponding to a reference size (corresponding  to a reference scale-length). In order to scale the intensity of the ISRF heating the diffuse dust in a galaxy, the parameters $SFR'$ and $old$ must be scaled to the reference size by comparing the scale-lengths (Eq.~D.3 in \citealt{2011A&A...527A.109P}): 
 
\begin{equation}
 SFR^{model}= SFR' \times \left( \frac{h_{s}^{ref}(B)}{h_{s}(B)} \right)^2
\end{equation} 

\begin{equation}
 old^{model}=old \times \left( \frac{h_{s}^{ref}(B)}{h_{s}(B)} \right)^2,
\end{equation}
where $h_{s}^{ref}(B)=5670\,pc$ and $h_{s}(B)$ are the reference B-band scale-length and the B-band scale-length of the galaxy under study, respectively. We note here that $SFR^{model}$ and $old^{model}$ are only internal parameters 
that allow us to interface with the library of models. Because of this 
all the results in this paper are presented only in terms of the $SFR$ and 
$old$, the real parameters of the galaxy under study.

An additional scaling is required to set the flux levels of the SEDs from the library to the observed SED of our galaxy. Thus, the SED that represents our galaxy, $F^{d}_{\lambda}$, is determined as (Eq.~D.2 in \citealt{2011A&A...527A.109P}):

\begin{equation}
F^{d}_{\lambda} = 
\left( \frac{h_{s}(B)}{h_{s}^{ref}(B)} \right)^2 \times F^{d,model}_{\lambda}(B/D,\tau_{B}^{f},SFR^{model},old^{model}) .
\end{equation}

We would like to stress that in these models the absolute flux level of the predicted dust SED is fixed by the input parameters.

The models of \citet{2011A&A...527A.109P} have been designed to provide the formalism for fitting the integrated emission of galaxies, since for most galaxies resolved information on the dust emission is not available. In particular dust emission SEDs of individual star-forming complexes cannot be derived in most cases. Consequently, the  model of \citet{2011A&A...527A.109P} provides an average template SED for the dust emission of the ensemble of the HII regions. This template has been empirically calibrated on data of a representative sample of prominent star forming complexes in our Galaxy, as fitted using the model of \citet{2008ApJS..176..438G}. By contrast, in this study of NGC~4214, we have resolved information on the central star forming complexes, and can therefore replace the average template with a detailed modelling of these complexes using directly the model of \citet{2008ApJS..176..438G}, while still retaining the general framework of the model of \citet{2011A&A...527A.109P}, when calculating the attenuation of stellar light in the clumpy component. This allows us to reach a self-consistent treatment of the diffuse and clumpy component, mediated by the fact that the two models used here have a common parameter, the $F$ factor.

\subsection{The model of Groves et al. (2008)}
     \label{HIImodels}

The model of \citet{2008ApJS..176..438G} describes the luminosity evolution of a star cluster of
mass \mcl, and incorporates the expansion of the HII
region and PDR due to the mechanical energy input of stars and SNe. The dust emission
from the HII region and the surrounding PDR is calculated from radiation transfer.
 The main input parameters  of the model are:

\begin{itemize}
 \item The metallicity of the star cluster, $Z$, in units of the solar metallicity $Z_{\sun}$ (\citealt{2005ASPC..336...25A}). The metallicities used in the model are restricted to
 the SMC and LMC metallicities. This limitation is introduced by the stellar population model (\citealt{2002MNRAS.337.1309S}).
 \item The age of the star cluster, \Age.
 \item The ambient pressure, \po, which controls the rate of the expansion of the HII region and PDR due to the mechanical energy input of stars and SNe,  and the size. Therefore \po\ determines the dust temperature for a given star cluster mass and age. 
 \item The compactness parameter, \Comp, which parametrizes the heating capacity of the star cluster and depends on \mcl\ and \po.
 \item The covering factor, \fcov, which represents the fraction of the surface of the HII region covered by the PDR. This is the same parameter as the F factor in the model of \citet{2011A&A...527A.109P}.
 \item The hydrogen column density of the PDR, $N_{\rm{HI}}^{\rm{PDR}}$.
 \end{itemize}

At a given age, the model self-consistently calculates the luminosity distribution of the star cluster 
and the radius of the inner edge of the PDR and computes the emergent SED for the dust in the  HII region without
considering the PDR, $F^{HII}_{\lambda}$, and 
for the dust from the HII region which is completely covered by a PDR, $F^{HII+PDR}_{\lambda}$.  The total emergent SED from the star forming region which is partially covered by the PDR, $F^{SF}_{\lambda}$, is then given by:

\begin{equation}
F^{SF}_{\lambda} = (1-f_{cov}) \times F^{HII}_{\lambda}
+ f_{cov} \times F^{HII+PDR}_{\lambda}
\end{equation}

%%%%%%%%%%%%%%%%%%%%
%%%% PARAMETERS %%%%
%%%%%%%%%%%%%%%%%%%%

\section{Constraints on the input parameters}
\label{sec:parameters}

The large amount of ancillary data covering a wide wavelength range, as well as the result of previous studies from the literature
allow us to determine or at least constrain  most of the input parameters for both models. 

\subsection{Input parameters for the model of Groves et al. 2008}
\label{sec:Groves-parameters}

The two main SF complexes NW and SE are formed by an ensemble of smaller HII regions. Ideally, all these smaller knots of SF should be modelled separately. However, IR/submm observations lack the angular resolution to perform such a detailed analysis. Due to this limitation, 
for both complexes we adopted parameter  ranges which are wide enough to enclose the values of their individual knots.
With the only exception of $N_{\rm{HI}}^{\rm{PDR}}$, for which we adopted the value 10$^{22}$ cm$^{-2}$, typical of molecular clouds in our own galaxy, the input parameters of the model of \citet{2008ApJS..176..438G} were observationally constrained:

\vspace*{0.2 cm}
\textit{\textbf{(i) Metallicity}:} The metallicities of both SF regions  have been measured by
\citet{1996ApJ...471..211K},  who found values of 12+log(O/H) = 8.17$\pm$ 0.02 for the NW and 8.27$\pm$ 0.02 for the SE region.
With the Solar abundance 12+log(O/H) = 8.66 \citep{2005ASPC..336...25A}, as used in \citet{2008ApJS..176..438G}, this gives
$Z= 0.32~Z_{\odot}$ for NW and $Z= 0.41~Z_{\odot}$ for SE.
We used a metallicity of Z=0.4~Z$_{\odot}$ and also tested Z=0.2~Z$_{\odot}$,  
which yielded a similar result (no template for Z=0.3~Z$_{\odot}$ is available in \citet{2008ApJS..176..438G}).

\vspace*{0.2 cm}

\textit{\textbf{(ii) Age}:} The age of the star clusters of NGC~4214 have been extensively studied by different authors.
\cite{1996ApJ...465..717L} presented HST \textit{Faint Object Spectrograph} (FOS) ultraviolet spectra covering the main star cluster of the NW complex. Using spectral synthesis modelling, \cite{1996ApJ...465..717L} reported for this stellar cluster an age of 4-5\,Myr.
\cite{1998A&A...329..409M} obtained optical long-slit spectra of NGC~4214 with the ISIS spectrograph of the William Herschel Telescope. 
Their bidimensional spectra covered both SF complexes. From the comparison of different observationally
determined parameters (equivalent width of \hbeta, Wolf-Rayet population, effective temperature and UV absorption lines) to
 synthesis models, they determined ages of  3$\pm$1\,Myr for both regions.
\cite{2000AJ....120.3007M} mapped the central part of NGC~4214 with the HST \textit{Wide Field Planetary Camera 2} (WFPC2). From the \halpha\ equivalent width, these authors determined an average age of 3.0-4.0\,Myr and 2.5-3.0\,Myr for the NW and SE regions, respectively.
More recently, \cite{2007AJ....133..932U} also used the WFPC2 data set to determine the ages of the star clusters by a likelihood-maximization technique from the photometric colours. They determined an age of 5\,Myr and of 2-4\,Myr for the star clusters within the NW and the SE complex, respectively.
Based on the dispersion found by these studies, we adopted ages of the NW and SE complexes of  3-5\,Myr and 2-4\,Myr, respectively. 

\vspace*{0.2 cm}

\textit{\textbf{(iii) Ambient pressure and Compactness}:} \po\ and \Comp\ were determined by comparing the expected and the observed radii of the individual HII regions as a function of the age. For a cluster of a given age it is possible to find different combinations of \po\ and \Comp\ that provide the observed radius of the expansion bubble. However, this degeneracy can be avoided when the mass of the star cluster is known, since both parameters \po\ and \Comp\ are related
 by the equation:

\begin{equation}
\label{eq:Mcl}
\rm log \left( \frac{M_{cl}}{M_{\odot}} \right ) = \frac{5}{3} \times log \left( \mathcal{C}  \right ) - \frac{2}{3} \times log \left( \frac{p_{0}/k}{cm^{-3} K} \right)
\end{equation}

From the values of the masses reported by \citet{2007AJ....133..932U} we obtained that for the NW complex the value of \po\ ranges from 7.0 to 8.0 and \Comp\ ranges from 5.5 to 6.5 depending on the HII region considered. For the case of the SE complex, we found that \po\ ranges from 6.0 to 7.0 and that \Comp\ ranges from 4.5 to 5.5. We used these intervals to constrain both parameters.

\vspace*{0.2 cm}

\textit{\textbf{(iv) Covering factor}:}  We assumed that the PDR consists of optically thick, homogeneously distributed clouds that surround the star cluster, leaving a fraction uncovered so that the light can escape unattenuated from these "holes".
To a good approximation, the intrinsic luminosity of the central star cluster is the sum of the observed luminosities of the stars, $L_{\rm star}$, and the luminosity re-emitted by the dust, $L_{\rm dust}$. The covering factor is then $f_{\rm cov}=L_{\rm dust}/(L_{\rm dust}+L_{\rm star})$. We obtained $L_{\rm dust}$ by integrating the best-fit model template (see Sect.~6) from 3\mi\ to 1.5\,mm. 
To calculate $L_{\rm star}$ we integrated the de-attenuated fluxes measured in our apertures for the SF regions from \textit{GALEX} FUV to IRAC 3\mi.
The exact value of the opacity in front of the SF regions due to
the diffuse dust layer is not known.
For this reason, we carried out two different estimates, the first based on  the lowest and the second on the highest  realistic opacity. 
i) We adopted the  foreground opacity of $\tau _{\rm V} = 0.35$ (\citealt{2007AJ....133..932U}) measured locally in front of the 
main stellar cluster of the NW region and assumed that this value is representative for both SF regions.
The stars in the NW region seem to have evacuated most of the surrounding material associated with the HII region 
so that this value is expected to be lower than the average disk opacity in this area.
 We obtained $f_{\rm cov}^{\rm NW} = 0.45$ and $f_{\rm cov}^{\rm SE} = 0.65$ for the NW and SE regions, respectively.
 ii) We used the value of  $\tau_{B}^{f}=2.0$ which is the best fit value derived from  our modelling of the diffuse emission (see Sect.~\ref{sec:RESULTS}). 
Assuming that the regions are in mid-plane and using the inclination angle $\rm i=44^{\circ}$ \citep{2008AJ....136.2563W}, the opacity in front of the SF complexes is  calculated as $\rm 0.5\times\tau_{B}^{f}/\cos(i)=1.4$. In this case, we obtained $f_{\rm cov}^{\rm NW} = 0.20$ and $f_{\rm cov}^{\rm SE} = 0.30$.
Thus, we obtain estimates for the covering factor of $f_{\rm cov}^{\rm NW} = 0.20-0.45$ and $f_{\rm cov}^{\rm SE} = 0.30 -0.65$ for both regions.

\subsection{Input parameters for the model of Popescu et al. (2011)}
\label{sec:Popescu-parameters} 

In the case of the model of \citet{2011A&A...527A.109P} the parameters that we could determine observationally are:\\

\textit{\textbf{(i) Bulge-to-disk ratio:}} Since no bulge is visible in the late-type galaxy NGC~4214 we set B/D = 0.\\

\textit{\textbf{(ii) Inclination angle:}} We fix $\rm i=44^\circ$ following \citet{2008AJ....136.2563W}.\\

\textit{\textbf{(iii) Scale-length}:} We determined the radial stellar scale-length in the B-band from a SLOAN g-band image. First, we removed the background of the image by calculating the average value of a wide set of small circular apertures placed strategically outside the disk of NGC~4214. Then, we removed the contamination from foreground stars and HII regions, including the central regions SE and NW. We used the IRAF task ellipse (STSDAS package) with steps of 20\arcsec\ to determine the isophotes. For all the isophotes beyond 100\arcsec\ from the centre, the PA and the ellipticity were fixed. Finally, we fitted an exponential function to the  surface brightness profile determined with the mean values in each isophotal ellipse. We considered the ellipses whose mean values and the corresponding uncertainties were above the background level.

\begin{figure}
\centering
\includegraphics[width=9cm]{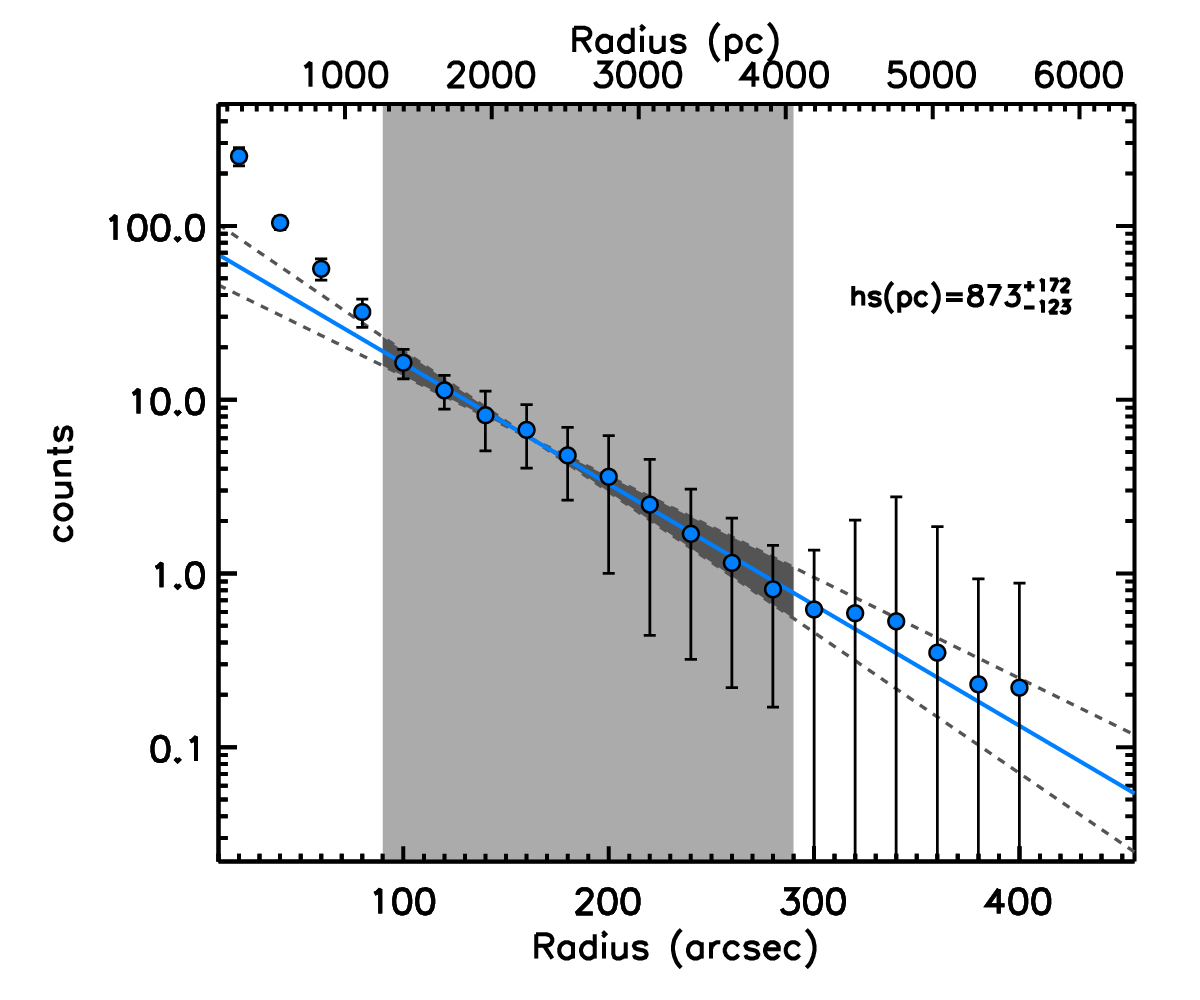}
\caption{B-band scale-length of the stellar disk of NGC~4214. The fit (solid
line) was achieved using the isophotes in the radial range marked by the
grey-shaded area. The uncertainty in the slope is represented by the dark
grey area and the dark grey dashed lines.}
\label{fig:SCALElength}
\end{figure}

We found that it is not possible to fit the profile for the whole disk with a single exponential function. For this reason we decided to calculate separately the scale-length in the inner and the outer part of the disk (see Fig. \ref{fig:SCALElength}). We visually determined the change of slope to take place at a radius $\sim$ 90\arcsec. The profile of the inner part of the disk (excluding the HII regions NW and SE) can be well fitted with a scale-length of $\sim$450 pc. For the outer part we determined a scale-length of 873$^{+172}_{-123}$ pc.\\

%----------------------------------------------
%----------------------------------------------

\textit{\textbf{(iv) Clumpiness}:} As a first estimate of the clumpiness factor $F$ we used the mean value from the upper and the lower 
limits obtained for the two SF regions, i. e., $\rm F=0.40\pm0.20$. This parameter is only used for the modelling  of the total integrated UV/optical/FIR emission described in Sect.~\ref{sec:fitTOTAL}. 
In Sect. 6.2 we will present an additional global  estimate of $F$.
 
\vspace*{0.2 cm}

\textit{\textbf{(v) old}:} We derived this parameter as $old = 0.009$ by assuming that all the luminosity in the J, H and K bands (see Table~\ref{tab:stellar_fluxes_DISK}) corresponds 
to the old stellar population. We integrated the luminosity of NGC~4214 in the J, H and K bands and derived $old$ as the ratio of this luminosity
and the integrated luminosity of the model galaxy (Table~E2 in \citealt{2011A&A...527A.109P}).

\begin{table}[h!]
%    \input{./Tables/STARS_NW_phot2tex}
%    \label{tab:stellar_fluxes_NW}
%    \caption{Region NW stellar flux densities of NGC~4214.}
%    \vspace{0.45cm}
%    \input{./Tables/STARS_SE_phot2tex}
%    \label{tab:stellar_fluxes_SE}
%    \caption{Region SE stellar flux densities of NGC~4214.}
%    \vspace{0.45cm}
   
%Table generated with dwarfSED.pro on Wed Aug 29 16:22:59 2012

%\begin{table*}
\centering
\begin{tabular}{c c c c} \hline \noalign{\medskip}
\textbf{BAND} & $\lambda_{0}$($\mu$m) & $F_{\rm DISK}$ (Jy) & Reference \\ \noalign{\smallskip} \hline \noalign{\medskip}
%%%%%%%%%%%%%%%%%%%%%%%%%%%%%%%%%%%%%%%%%%%%%%%%%%%%%%%%%%%%%%%%%%%%%%%%%%%%%%%%%%%
FUV & 0.154 & 0.072 $\pm$ 0.007 & 1 \vspace{0.15cm} \\ 
NUV & 0.232 & 0.091 $\pm$ 0.009 & 1 \vspace{0.15cm} \\ 
B   & 0.445 & 0.354 $\pm$ 0.056 & 2 \vspace{0.15cm} \\ 
V   & 0.551 & 0.446 $\pm$ 0.064 & 2 \vspace{0.15cm} \\ 
J   & 1.220 & 0.520 $\pm$ 0.014 & 3 \vspace{0.15cm} \\ 
H   & 1.630 & 0.614 $\pm$ 0.023 & 3 \vspace{0.15cm} \\ 
K   & 2.200 & 0.458 $\pm$ 0.022 & 3 \vspace{0.15cm} \\ 
%%%%%%%%%%%%%%%%%%%%%%%%%%%%%%%%%%%%%%%%%%%%%%%%%%%%%%%%%%%%%%%%%%%%%%%%%%%%%%%%%%%
\hline \\
\end{tabular}
%\label{DustDISK}
%\caption{Region DISK and diffuse photometric points.}
%\end{table*}

   \caption{Total stellar flux densities of NGC~4214. (1) This work, (2) \citet{1991rc3..book.....D}, (3) \citet{2003AJ....125..525J}.\label{tab:stellar_fluxes_DISK}}
\end{table}

%%%%%%%%%%%%%%%%%
%%%% RESULTS %%%%
%%%%%%%%%%%%%%%%%

\section{Results}
\label{sec:RESULTS}

In a first step we separately fitted the emission from the SF regions and from diffuse dust. In a second step we self-consistently combined the emission from the SF regions with that from the diffuse dust to fit the total emission of the galaxy.

The emission from the NW and SE regions have been directly  measured from the maps as described in Section~\ref{photometry}. The diffuse emission has been calculated as the difference between the total emission and the sum of the emission from the two SF complexes. For the data points where no direct measurement was available, the best-fit model value (see below) was taken. The values for the diffuse dust emission are listed in Table~\ref{tab:fluxes_diffuse}. %\ref{tab:fluxes_diffuse}.

In order to take into account the emission from smaller and less intense  SF regions in the disk of NGC~4214, we estimated their contribution from  the H$\alpha$ emission of the ten brightest secondary SF regions  which corresponds to 18\% of the
\halpha\ emission of SE+NW.
We assume that the shape of the dust SED from these smaller SF regions is the same as the sum of the SEDs of SE+NW, and  we  
include their contribution in the calculation of the lower error range of the fluxes of the diffuse dust component in each band (see Table 3). 
The smaller HII regions only have a noticeable effect on the emission at 24\mi\ and 70\mi.

\subsection{Best fits for the SF regions}
\label{sec:fitHII}

\begin{figure*}
 \centering
 \includegraphics[width=0.85\textwidth]{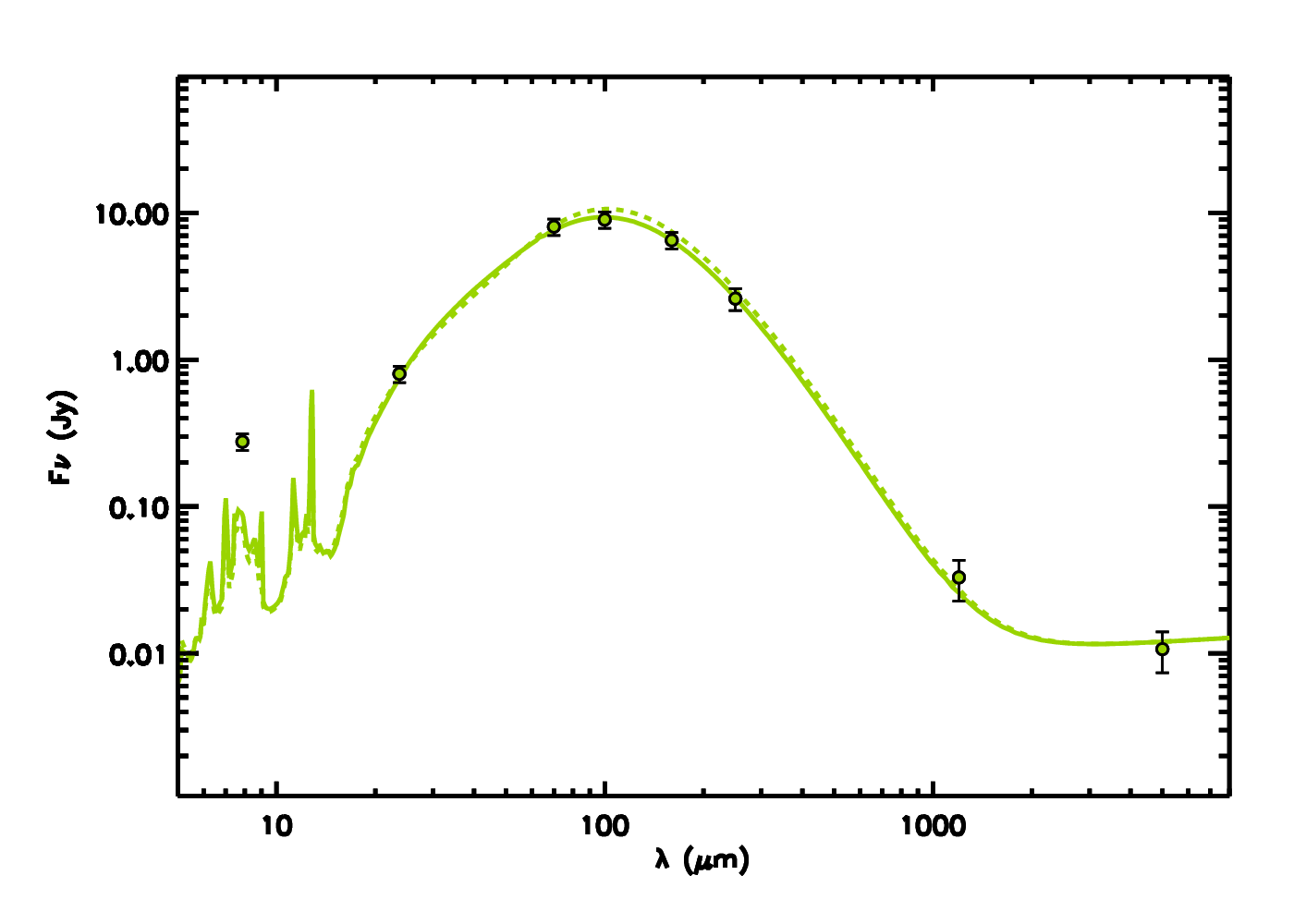}
 \caption{Best-fit models to the NW complex for Z=0.4\,\Zsun\ (solid line) and Z=0.2\,\Zsun\ (dashed line). The values of the reduced $\chi^2$ (neglecting the IRAC\,8\mi\ point) are 0.25 and 0.75 for Z=0.4 and 0.2\,\Zsun, respectively.}
 \label{fig:fitNW}
%  \end{figure*}
% 
%  \begin{figure*}
 \centering
 \includegraphics[width=0.85\textwidth]{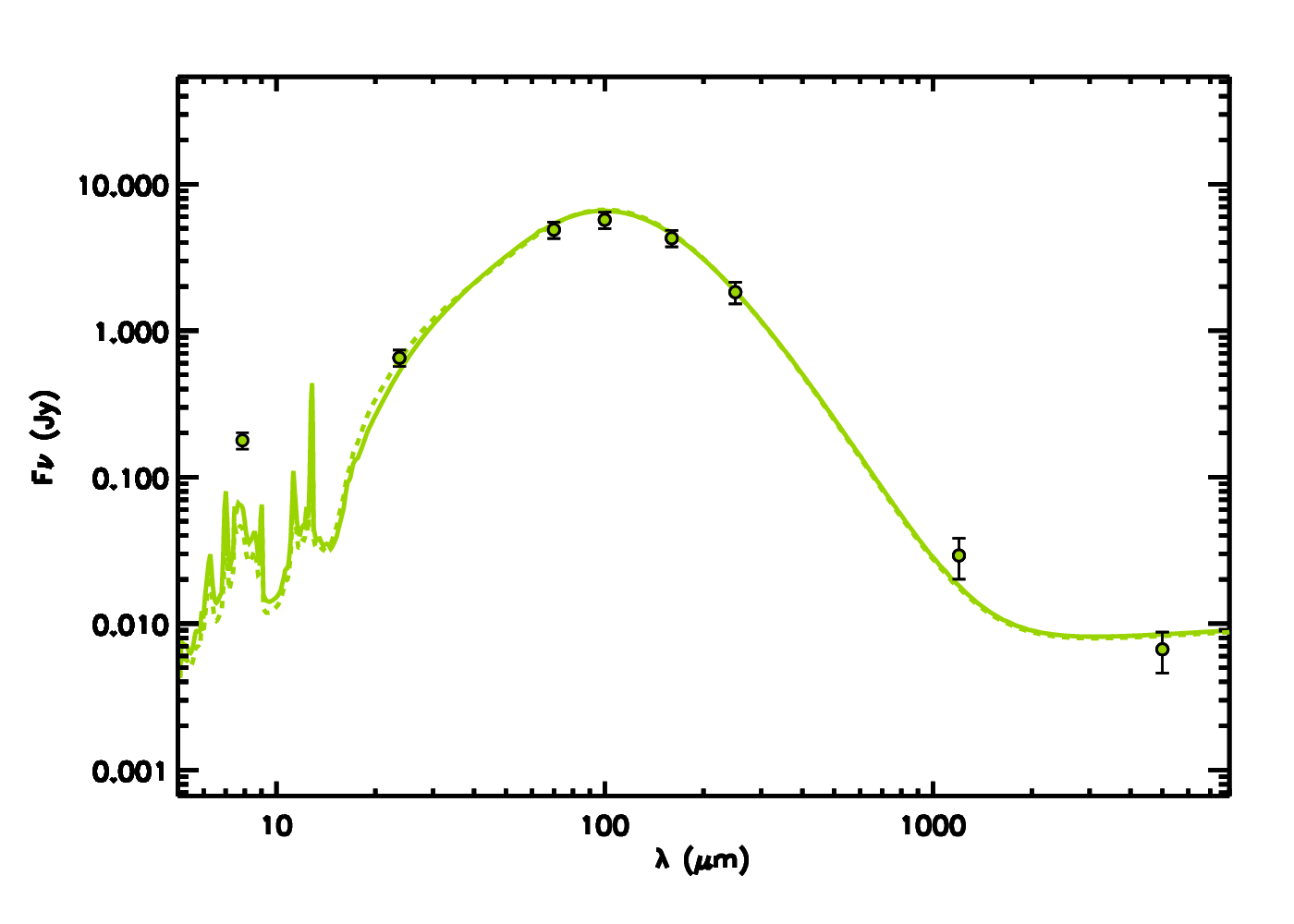}
 \caption{Best-fit model to the SE complex for Z=0.4\,\Zsun\ (solid line) and Z=0.2\,\Zsun\ (dashed line). The values of the reduced $\chi^2$ (neglecting the IRAC\,8\mi\ point) are 1.17 and 0.89 for Z=0.4 and 0.2\,\Zsun, respectively.}
 \label{fig:fitSE}
\end{figure*}

The best-fit models for the NW and SE complexes are shown in Figures \ref{fig:fitNW} and \ref{fig:fitSE}, respectively.

We obtained a good fit for all our observed photometric points longwards of 10\mi. For a metallicity of Z=0.4\,\Zsun\ the best-fit models of the two complexes correspond to the parameters \Age$\rm=4.0\,Myr$, \Comp$\rm=5.0$, \po$\rm=8$ and \fcov$=0.30$ for both regions. With the exception of the \Comp\ of the NW region and  \po\ of the SE region, all the parameters fall within our parameter 
 ranges constrained from the observations (see Sect.~\ref{sec:Groves-parameters}).
For a metallicity of Z=0.2\,\Zsun\ the best fit parameters are
in the case of the NW region \Age$\rm=5.0\,Myr$, \Comp$\rm=5.0$, \po$\rm=7$ and \fcov$=0.30$ and in the case of the SE region \Age$\rm=3.5\,Myr$, \Comp$\rm=4.5$, \po$\rm=8$ and \fcov$=0.60$. The value for \Comp\ in NW is slightly lower, and \po\ in SE slightly higher, but the rest of the parameters are
within the observed ranges.

We note that the IRAC\,8\,\mi\ data point was excluded from our fitting procedure. The reason is that both models considerably underestimate the emission at 8~\mi. Specifically, from our best-fit models for Z=0.2\,\Zsun\ we obtained that the observed fluxes are larger than the model values by a factor of 3.8 and 4.0 for NW and SE, respectively. In the case of the best-fit models of Z=0.4\,\Zsun\ the discrepancy decreases to a factor of 3.1 and 2.9 for NW and SE, respectively. A complete discussion of the IRAC\,8\,\mi\ discrepancy will be given in Sect.~\ref{sec:Discussion}.

\subsection{Best fit for the diffuse emission}
\label{sec:fitDIFF}

%----------------------------------------------------------
\begin{figure*}
\centering
\includegraphics[width=0.99\textwidth]{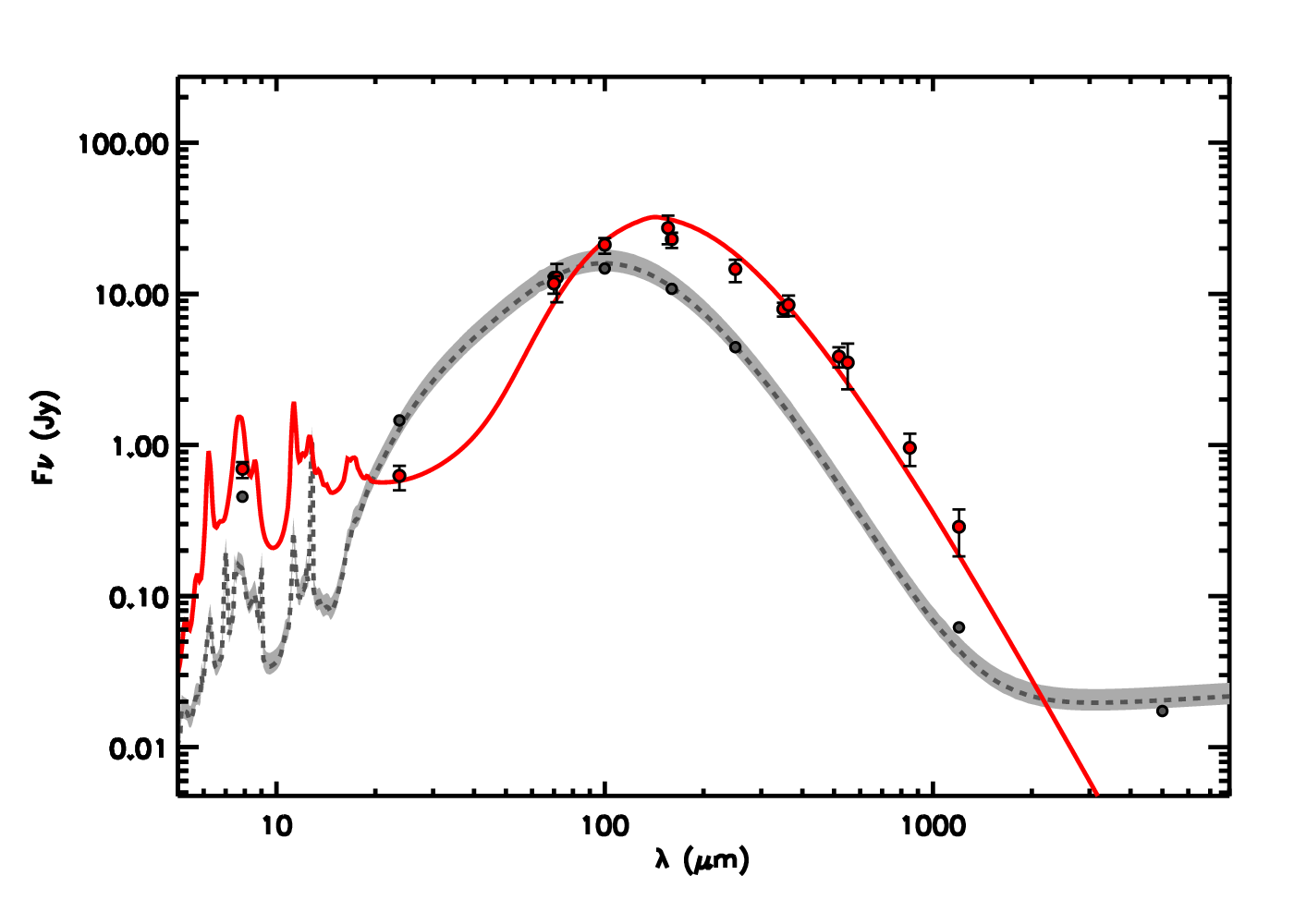}
\caption{Fit to the diffuse emission of NGC~4214 keeping fixed $old =
0.009$ and leaving \taubf\ and $SFR'$ as free parameters. The red solid
line represents the best fit to the MIR/submm SED for the measured value
of the scale-length \hs{873}, obtained for \taubf$\rm=2.0$, and
\SFR{\prime}{}{0.059}. The dark grey dashed line corresponds to the sum
of the best fit models of the HII regions presented in Figs.
\ref{fig:fitNW} and \ref{fig:fitSE} and the grey-filled circles
correspond to the sum of the photometric points of NW and SE regions. The
light grey area represents the uncertainty in the estimation of the total
emission of the HII regions.
The value of the reduced $\chi^2$ (neglecting the IRAC\,8\mi\ and
MAMBO points) is 1.82.}
\label{fig:fitDIFF}
\end{figure*}
%----------------------------------------------------------

We searched the library of the diffuse dust SEDs of
\citet{2011A&A...527A.109P} for the best fit to the data in the MIR/submm
range, leaving \taubf\ and \sfrprime\ as free parameters and keeping $old$
fixed to the value 0.009. Fig. \ref{fig:fitDIFF} shows the best fit
obtained for our measured value \hs{873} (solid red line), where we
determined as best-fit values $\tau_{B}^{f}=2$ and \SFR{\prime}{}{0.059}.
Neither the data point at 8\mi\ micron nor at 1.2\,mm were taken into account in the $\chi^2$ fitting procedure, the latter due to the observational limitations discussed in Sect.~\ref{sec:APERT}.

The model fits the data points in general very well.  The largest
discrepancy occurs at 8~\mi\ where the model overpredicts the observations
by a factor of 1.9. We will discuss this further in
Sect.~\ref{sec:Discussion}. At 160~\mi\  there is a discrepancy with the
PACS value which is overpredicted by 34\% by the model, but  the model
prediction agrees within the errors with the MIPS data point at the same
wavelength. The model overpredicts the SPIRE 250~\mi\ point by 24\% and
underpredicts the 850~\mi\ point by 34\%. For the other data points the
model predictions agree within the error bars.

It is instructive to compare this value of \sfrprime\, derived from
the fit to the  diffuse component of the dust emission, with the
value for \sfrprime\, derived from our measurement of the
spatially integrated flux density in the UV-to-blue band and which we will  call, for clarity,
$SFR{^\prime}_{\rm UV}$ in the following.
Applying the attenuation
corrections due to diffuse dust, as
tabulated in \citet{2011A&A...527A.109P} for $\rm
\tau_{B}^{f}=2$ and an inclination of $\rm 44^{\circ}$  we derive a value of 
$L_{\rm UV} = 3.70 \times 10^{35} $ W. We then determine  $SFR{^\prime}_{\rm UV}$ 
following Eq. 17 in \citet{2011A&A...527A.109P} as 
$SFR{^\prime}_{\rm UV} =
L_{\rm UV}/L^{young}_{unit,UV}$, where $L^{young}_{unit,UV}$ is the
normalization factor. We obtain \SFR{\prime}{UV}{0.165}. 
This is a factor of 2.8 times  higher than the value
of \sfrprime\ derived from the fitting of the diffuse dust emission. The most
straightforward explanation for this discrepancy would be that the true
value for the  disk scale length $h_{\rm s}$ is higher than the adopted
of 873\,pc. This would favour a disk with lower dust surface
densities, leading to a higher value of \sfrprime\ needed to account for
the observed  amplitude and colour of the diffuse dust emission. If we
adopt for the scale-length the maximum allowed by the B-band images, 
$h_{\rm s} = 1045\,pc$, we obtain a best fit to the diffuse dust emission SED of
$\rm \tau_{B}^{f}=1.2$ and \SFR{\prime}{}{0.088}. 
At the same time, this reduction in $\rm \tau_{B}^{f}$ means that 
the deattenuated, integrated UV luminosity, $L_{\rm UV}$, is now lower,
yielding \SFR{\prime}{UV}{0.147} which is only a factor of 1.67 greater than the
value of \sfrprime\ derived from the diffuse dust emission. A complete
discussion of this discrepancy will be given in
Sect.~\ref{sec:Discussion}.

As a further consistency check we can use the value of \sfrprime\
derived from the diffuse dust emission to make a different, global
estimate of the fraction of the UV radiation escaping from the SF regions,
\fesc, and the corresponding clumpiness factor, $F$, and compare it with
the  corresponding value already derived in Sect.~\ref{sec:parameters}
from analysis of the direct and dust-re-radiated UV light seen from the
spatially resolved SF regions. For this, we have to  assume that the
effective star formation rate powering the diffuse dust, \sfrprime, is due
to radiation escaping from the regions NW and SE. The total effective SFR
produced by NW+SE is then \sfrprime+$SFR_{\rm loc}$ where $SFR_{\rm loc}$
is the effective star-formation rate needed to power the dust emission
from the SF regions. We calculate $SFR_{\rm loc}$ by integrating the dust 
SED from the SF regions over the entire wavelength range, obtaining
$L_{\rm dust}$,  and assuming that this luminosity is equivalent to the
same amount of UV radiation absorbed by the dust locally. We then
determine $SFR_{\rm loc}$ in an equivalent way as described for $SFR^\prime_{\rm UV}$ above, as $SFR_{\rm loc} =
L_{\rm dust}/L^{young}_{unit,UV}$.
%where $L^{young}_{unit,UV}$ is the
%normalization factor \citep[see Eq.~17 in][]{2011A&A...527A.109P}. 
We obtain  $L_{\rm dust} = 9.06 \times 10^{34}$  W, which  gives \SFR{}{\rm
loc}{0.040}. With \SFR{\prime}{}{0.059}, derived from our best fit for the
diffuse emission, we then obtain \fesc = $SFR'/(SFR'+SFR_{\rm loc})$ =
60\% of the UV radiation of the SF region is required to escape from the
SF complexes in order to heat the diffuse dust. This escape fraction
corresponds to a \fcov\ $=F=1-0.60=0.40$,  which is the same value derived
in Sect.~\ref{sec:parameters}. We thus adopt in the following $F =
0.40\pm0.20$ (where the error comes from the range of \fcov\ derived in
Sect.~\ref{sec:Groves-parameters}). From the best-fit value of \sfrprime\
and $F=0.40$ we can estimate the value of the total star formation rate of
NGC~4214 as $SFR = SFR^\prime/(1-F) = 0.059/(1-0.40) = 0.098$ \Msun
yr$^{-1}$.

%%%%%%%%%%%%%%%%%%%%%%%%%%%%%%%%%%%%%%
\subsection{Best fit for the total emission}
\label{sec:fitTOTAL}
 
\begin{figure*}
  \centering
  \includegraphics[width=0.99\textwidth]{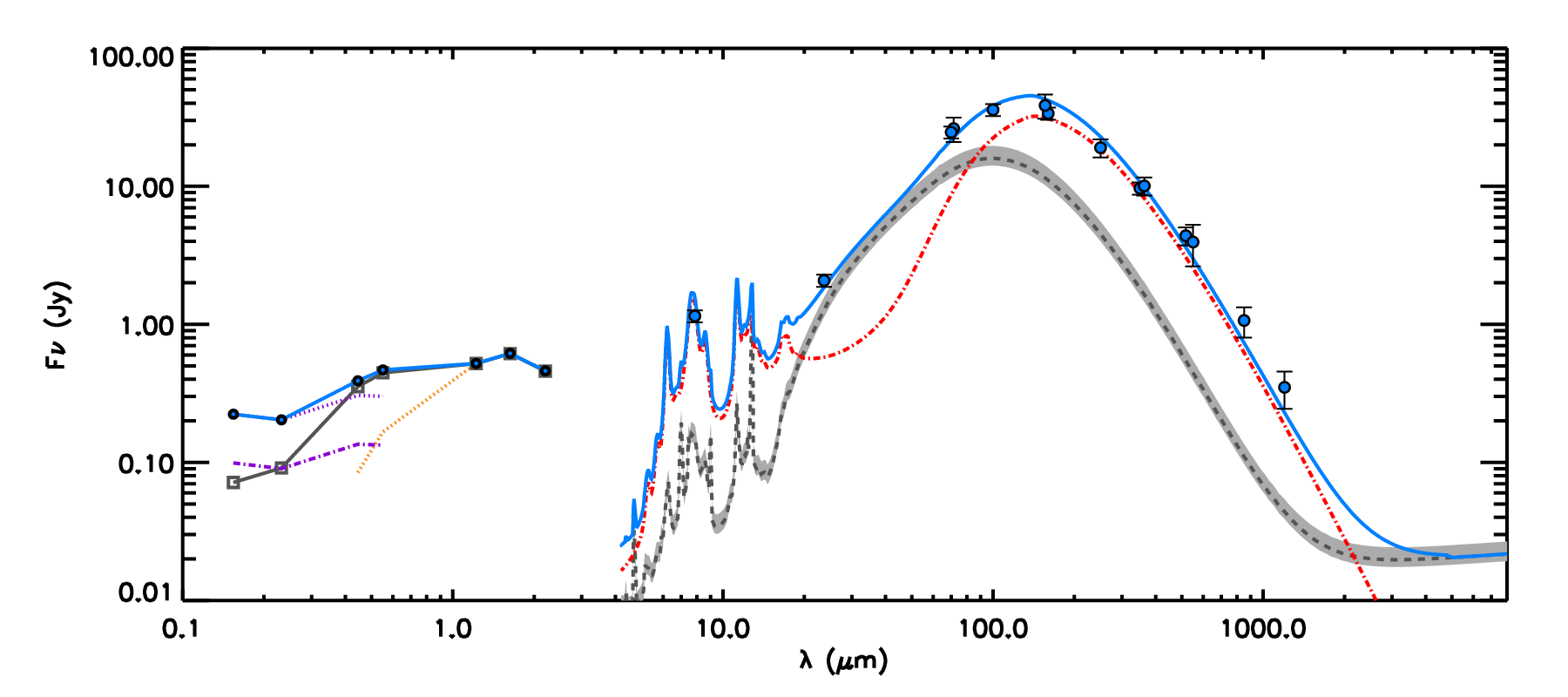}
  \caption{SED for the total emission of NGC~4214. On the MIR/FIR/submm part we show the best-fit solution for the total emission (blue solid line) obtained as the sum of the best fit models of the HII regions (grey dashed line) from Figs.~\ref{fig:fitNW} and \ref{fig:fitSE} and the best fit model to the diffuse emission (red dashed-dotted line) from Fig.~\ref{fig:fitDIFF}, obtained for
   \taubf$\rm=2.0$ and \SFR{}{}{0.098}.
  On the UV/optical/NIR part, the grey squares and grey solid line represent the observational data, the blue-filled circles and the blue solid 
  line are the intrinsic fluxes derived
  from deattenuation of the observed data points as described in Sect.~\ref{sec:fitTOTAL}. 
    % taking into account the attenuation from diffuse dust and from dust in HII regions. 
  The orange dotted line corresponds to the intrinsic emission of the old stellar population (see Sect.~5.2 for details). The dotted purple line shows the intrinsic emission from the young stellar population from the UV to the optical wavelength range. The emission of the young stellar population in the B and V band was obtained after subtraction the contribution of the old stellar component (orange dotted line).  
  %  , which goes from the intrinsic UV data points to the optical and was
%obtained in the B and V bands by   subtraction of the old stellar component from the intrinsic emission at these wavelengths. 
The purple dashed-dotted line is the scaled emission of the young stellar population by the factor needed to predict the same $SFR$ as derived from the dust SED modelling.}
  \label{fig:fitTot}
 \end{figure*}

On the MIR/FIR/submm part of Fig.~\ref{fig:fitTot} we show the fit to the total dust emission, which is the sum of the SEDs from the SF regions and the SED from the diffuse dust obtained in the previous sections. The entire dust SED from 8\mi\ to 850\mi\ can be well fitted. 

On the UV/optical/NIR part of Fig. \ref{fig:fitTot} we show the observed UV-optical SED (grey squares connected by a solid line).
Following Eq. C.12 of \citet{2011A&A...527A.109P}, the UV/optical SED of the young stellar disk of NGC~4214 was derredened using the composite attenuation,  $\Delta m{_\lambda}$, which is for the case of $old\sim0$:

\begin{equation}
    \label{eq:atten}
   \Delta m{_\lambda}=-2.5 \log \left( 1-F\,f_{\lambda} \right) + \Delta m{_\lambda}^{tdisk}
\end{equation}
where the first part takes into account the attenuation in the SF regions and $\Delta m{_\lambda}^{tdisk}$  the attenuation of the diffuse component. The wavelength dependence of the escape fraction, $f_{\lambda}$, is tabulated in Table A.1 in \citet{2004A&A...419..821T}. For the old stellar population we used the attenuation correction derived for the old stellar component in the model of \citet{2011A&A...527A.109P} $\Delta m{_\lambda}=\Delta m{_\lambda}^{disk}$. For both stellar populations we used the value of \taubf$\rm=2.0$ obtained from our best-fit model and an inclination angle of  $i = 44^{\circ}$ (\citealt{2008AJ....136.2563W}). The SED derived in this way represents the total intrinsic stellar SED of NGC~4214 and is shown in the blue line.
 
We now calculate, in a similar way as in Sect. 6.2 for \sfrprime , from the intrinsic UV-to-blue luminosity the predicted value  of the SFR, by applying Eq.~17 from \citet{2011A&A...527A.109P}, $SFR=L_{UV}^{tdisk}/L_{unit,UV}^{young}$. Using $F=0.40$ in Eq.~\ref{eq:atten}, we derive \SFR{}{}{0.22}, which is a factor of 2.24 higher than the value derived from the fitting of the dust SED (\SFR{}{}{0.098}).
For $F=0.20$ and $F=0.60$, the highest and lowest values suggested by our data, the corresponding ratios are 2.51 and 2.18, respectively. If we adopt for the scale-length the maximum value allowed by the data \hs{1045}, we obtain the best fit for $\tau_{B}^{f}=1.2$ and \SFR{\prime}{}{0.088}, yielding $SFR = SFR^\prime/(1-F) = 0.088/(1-0.40) = 0.14$ \Msun yr$^{-1}$. In this case, the discrepancy with the value derived from the derredened UV-optical data, \SFR{}{}{0.188}, decreases to a factor of 1.34 for $F = 0.40$ (1.50 and 1.27 for the $F=0.20$ and $F=0.60$, respectively).

A complete discussion of this discrepancy will be given in Sect.~\ref{sec:Discussion}.
 
%---------------------------------------------

\subsection{Gas-to-dust mass ratio}
\label{sec:DUSTtoGAS}

The total diffuse dust mass can be calculated from Eq.~44 of \citet{2011A&A...527A.109P}
($\rm M^{diff}_{dust} = \tau_B \times h_s^2 \times 0.99212$ pc$^{-2} M_\odot$).
With our values of $h_{\rm s} = 873\,$pc  and $\tau_B = 2$  we obtain
 $\rm M^{diff}_{dust} = 1.5 \times 10^6$ $M_\odot$.  
This value is quite robust again changes in 
parameters: Choosing the largest permitted scale-length, $h_{\rm s} = 1045\,$pc and
the corresponding best fit for the opacity, $\tau_B = 1.2$,
we obtain  $\rm M^{diff}_{dust} = 1.3 \times 10^6$ $M_\odot$, only  13\% lower. 
From the modelling of the HII complexes for Z=0.4\,\Zsun, we derived dust masses of
$\rm M^{NW}_{dust}=0.79\times10^5$\,\Msun\ and
$\rm M^{SE}_{dust}=0.42\times10^5$\,\Msun\ for the NW and SE complexes, respectively.
% For the case of Z=0.2\,\Zsun\ we derived
% $\rm M^{NW}_{dust}=0.85\times10^5$\,\Msun\ and
% $\rm M^{SE}_{dust}=0.52\times10^5$\,\Msun.
We therefore derived for NGC~4214 a total mass of dust $\rm M^{total}_{dust}=M^{diff}_{dust}+M^{NW}_{dust}+M^{SE}_{dust}=1.62\times10^6$\,\Msun. 

The total gas mass can be derived from  $M_{\rm HI} = 4.1 \times 10^8$ $M_\odot$
(\citealt{2008AJ....136.2563W}) and $M_{\rm H2} = 5.1 \times 10^6$ $M_\odot$ (\citealt{2001AJ....121..727W}, obtained with a Galactic conversion factor) and gives, taking into account a helium fraction of 1.36,
$M_{\rm gas} = 5.66 \times 10^8$ $M_\odot$. The total gas-to-dust mass ratio is then $\rm G_{dust}=350$.

The Galactic CO-to-H$_2$ conversion factor (X-factor, adopted here as
N(H$_2$)/$I_{\rm CO} = 2\times10^{20}$ cm$^{-2}$ (K \kms)$^{-1}$),
most likely  severely underestimates  the total molecular gas mass in
NGC~4214. % which would only be about 1\% of the total gas mass based on this X-factor. 
An  indication for this is e.g.  the very high CII/CO ratio measured  in the central region \citep{2010A&A...518L..57C} which shows that 
a large fraction of the CO is photo-dissociated due to the high radiation field and low dust shielding.
Even though the X-factor has been notoriously difficult to determine, progress has been made in
recent years mainly due to 
the possibility to derive dust masses with an increasingly better precision. This allows, 
together with HI and CO measurements 
and the assumption of a constant gas-to-dust mass ratio, to 
 derive the ratio between total molecular gas mass and CO emission.
\citet{1997A&A...328..471I}  has first used this method and derived
for NGC~4214 a 15-30 times higher X-factor than  the Galactic value.
\citet{2011ApJ...737...12L} used spatially resolved data for
the dust masses, derived  from  fits to HERSCHEL and SPITZER data, and the  gas mass from HI and CO measurements
for a small sample of nearby galaxies (M~31, M~33, LMC, SMC and NGC~6822) .
%Assuming that the gas-to-dust ratio and the X-factor are constant over the area considered they estimated both
%parameters by minimizing the scatter in the data. 
They found evidence for a strong increase of the X-factor
below metallicities of 12-log(O/H) = 8.2-8.4, most likely due to the dissociation of CO and the creation of extended
layers of CO-free H$_2$.  \citet{2011ApJ...737...12L}  find for NGC~6822,
which has a similar  metallicity as NGC~4214 (12+log(O/H) = 8.2) an X-factor of 4-5 times the Galactic value.
For the same galaxy, NGC~6822, and a similar method, \citet{2010A&A...512A..68G}  derived a  higher X-factor 
(20 times the Galactic value).
Taking this range of estimates into account, we adopt  an X-factor 10 times the
Galactic value as a reasonable estimate and derive for this case a  gas-to-dust mass ratio  of  387. 

If we assume that the gas-to-dust mass ratio scales linearly with metallicity (which means that the fraction
of metals incorporated in the dust is constant) we expect, based on the solar value of about
150, a value of $\rm G_{dust}$ between 375 (for Z = 0.4~Z$_\odot$) to 500 (for Z = 0.3~Z$_\odot$).

Thus, the observed gas-to-dust ratio is very close to the expected value in the case of Z = 0.4~Z$_\odot$,
and a factor 1.3 lower for Z = 0.3~Z$_\odot$. Even though we consider this discrepancy small,
it could indicate that there is a still larger amount of "dark" molecular gas,
not probed by the CO. We would need a conversion factor 35 times the Galactic value in order
 to obtain a gas-to-dust mass ratio of 500. 
An alternative possibility is that the dust properties in NGC 4214 are different with a higher dust extinction
coefficient in the submillimeter wavelength range. This could be  produced e.g. by amorphous graphite as suggested by \citet{2011A&A...536A..88G} for the case of the LMC.

%%%%%%%%%%%%%%%%%%%%
%%%% DISCUSSION %%%%
%%%%%%%%%%%%%%%%%%%%

\section{Discussion}
\label{sec:Discussion}

The analysis presented in this paper enables us to draw conclusions
about the physical properties of the dwarf galaxy under study. The
way this can be achieved is by confirming or rejecting the basic
ingredients of the models used to fit the data, from the consistency,
% or otherwise,
between model predictions and data. Since our models
are radiative transfer models, they contain a wealth of information
regarding the distribution of stars and dust in galaxies, the
clumpiness of the ISM, the dust opacity of the dust
clouds, the contributions of the different phases of the ISM, and the
optical properties of the dust grains, including the relative
abundance of PAH molecules.

Although we can fit the FIR SED of this galaxy, the main discrepancy
is that the UV emission is underpredicted with respect to our
corresponding predictions for attenuation of UV light. We discuss
possible causes for this discrepancy, as well as consequences for the
derived physical parameters of this galaxy. We also discuss the
underprediction of the 8\mi\ emission in the two central
star-forming regions, and its implications.

\subsection{Emission from PAHs at 8\mi}

We found that the observed emission at 8~\mi\ is overpredicted for the diffuse emission, and severely underpredicted for the emission from the HII regions.
One reason for this could be, as we discussed in Sect.~\ref{sec:APERT},  that part of the 8\mi\ flux seen in the apertures, although powered by UV light from the star cluster, is actually likely to originate in the diffuse dust layer beyond the PDR+HII region considered by the \citet{2008ApJS..176..438G} model. 
Although it is impossible to quantify the importance of this effect, we do not think that it is entirely responsible for the discrepancies, because most of the 8~\mi\ emission is clearly associated with the SF regions, as shown by the close correspondance between
the 8~\mi\  and  \halpha\ distribution (e.g. in  the shell structure in the  NW region.)

The model of \citet{2011A&A...527A.109P} assumes Galactic dust properties and a fraction of PAHs in dust which is appropriate for our Galaxy. The PAH fraction in NGC~4214 might be lower given the lower metallicity (e.g. a lower PAH fraction has been found in the diffuse ISM in the SMC by \citealt{2010ApJ...715..701S}). If this is the case, we can quantitatively understand the discrepancy.

A possible reason for the excess of 8\mi\ emission in the HII regions may be that the model of \citet{2008ApJS..176..438G} overpredicts the PAH destruction for this particular object. In the models of \citet{2008ApJS..176..438G} the PAH abundance is fixed to be proportional to the gas metallicity. In addition to this, it includes a  parameter which controls the destruction of the PAHs in intense radiation fields. Thus, this model considers PAH destruction and assumes a lower fraction of PAHs in dust in the SF regions.

The observations of \citet{2008ApJ...678..804E} indeed suggest that NGC~4214 might be a galaxy with an unusually high 8~\mi\  emission for a given radiation field. They found a good correlation between the equivalent width at 8\mi, EW(8\mi), and the ionization parameter for a set of starburst galaxies. The relation has also been confirmed for HII regions in M101 \citep{2008ApJ...682..336G}. NGC~4214, one of the starburst galaxies in the \citet{2008ApJ...678..804E} sample, deviates significantly from this correlation: it exhibits an EW(8\mi) a factor of 3 higher than the one corresponding to the ionization parameter. The spectra used in this analysis are not restricted to the SF regions and thus might contain emission from the diffuse medium (\citealt{2008ApJ...678..804E} do not quantify this)  and therefore we cannot rigorously compare this result to our findings for the SF complexes. However, since the lines from the ionized atoms and the PAH emission come to a large extent from HII regions and their PDRs (see our Tabs. 1-3)  the result of  \citet{2008ApJ...678..804E} suggests that the 8~\mi\ emission in the SF regions in NGC~4214 might indeed be high.

A final possibility to explain the high 8~\mi\ emission from the SF region is the ratio between neutral and ionized PAHs. The model of \citet{2008ApJS..176..438G} uses templates fitted to the starburst galaxies NGC~4676 and NGC~7252 with a corresponding ratio between neutral and ionized PAHs. This ratio might not be entirely appropriate for the low-metallicity galaxy NGC~4214. Ionized PAHs emit about a factor of 10 more energy in the 6-9\mi\ range than at 11-12~\mi, whereas neutral PAHs emit about the same amount in both ranges \citep{2007ApJ...657..810D}. Thus, a higher ionized PAH fraction could boost the predicted emission at 8~\mi\  at the expense of the emission at 11-12~\mi. If we assume as an extreme case that the entire emission at 11-12~\mi\ is transferred to the 8~\mi\ band, we can increase the emission for our model spectrum at 8~\mi\ by a factor of $\sim$ 2. Thus, we could indeed improve the agreement between data and model, although some discrepancy would remain.

\subsection{The UV emission illuminating the diffuse dust}

In Section 6 we found that there is tension between the observed surface brightness of the disk in the submm and the observed flux of the galaxy in the non-ionising UV in the sense that, although the FIR SED can be fitted, the UV emission is underpredicted by a factor 
$\sim$ 2-3 with respect to our corresponding predictions for attenuation. Here we discuss the plausibility of such an escape fraction and study further possibilities to explain the differences between model and data.

\subsubsection{Escape of unattenuated UV radiation from NGC~4214}

%{\bf In order to match data and model we require that about 30-70\% of the observed UV-optical luminosity does not contribute to the dust heating, and instead leaves the galaxy unattenuated.}

In starburst galaxies, it has been previously  observed that a fraction of the UV luminosity can escape from the galaxies without heating the dust, mainly due to the porosity of the ISM created by the energy input of massive stars and SNe which in the extreme case leads to (super)bubbles and galactic winds. \citet{2007ApJ...661..801O} showed that  starburst galaxies present a low diffuse H$\alpha$ emission consistent with an expected escape fraction of ionizing photons of 25\% from the galaxy. 

Other studies show that the escape of ionizing radiation (and therefore non-ionizing as well) could be much lower: \citet{1995ApJ...454L..19L} pointed out that 3\% of the intrinsic LyC photons escape from a set of starburst galaxies; \citet{2009ApJS..181..272G} found an even lower ($<$1\%) fraction for local starburst galaxies, and \citet{2007ApJ...668...62S} found a low escape fraction for low (subsolar) metallicity starburst galaxies ($<$8\%).

However, the detection of the escaping photons might be determined by 
the geometry and orientation of the observations if holes and open 
channels are the medium that the radiation uses to escape from the 
galaxy (\citealt{2002MNRAS.337.1299C}, \citealt{2011ApJ...741L..17Z}). Besides, the escape fraction can be 
affected by the internal properties of galaxies, depending sensitively 
on the covering factor of clumps, and the density of the clumped and 
interclumped medium (\citealt{2011ApJ...731...20F}). 

Conditions that are favorable for an escape are likely to be given in starburst dwarf galaxies which generally show large bubbles and shell structures \citep{1998ApJ...506..222M} -- similar to what we are seeing in the center of NGC~4214 -- that would make the ISM more porous and prone to enable the UV radiation to escape. The simulations of \citet{2009ApJ...693..984W} showed that an  escape fraction as high as 25-80\% can  be expected  due to the irregular morphology of the dwarf galaxies with a clumpy ISM.  There are observational confirmations of such an escape. 
 \citet{2012arXiv1209.0804C} found evidence for a escape of 80\% of the non-ionizing UV-radiation from the low metallicity Haro~11 and
 \citet{2006A&A...448..513B}   an escape fraction of the Lyman continuum radiation of 4-10\% in the same 
 object.  \citet{2011ApJ...741L..17Z} have detected an ionized cone above a shell structure in the starburst galaxy NGC 5253 suggesting that ionizing radiation could be escaping from this galaxy. In the presence of such channels we would expect that not only ionizing, but also non-ionizing radiation escape from the galaxies without being affected by dust.

The geometry of the central SF regions, especially in the case of the NW region, suggests that bubbles have been already created in NGC~4214. We can  make an estimate of how much such a process could affect NGC~4214 by assuming that the entire UV luminosity from these regions leaves the galaxy unattenuated. The UV luminosity of the two main HII regions represent 20\% of the total UV luminosity of NGC~4214. Therefore, about 20\% of the total UV luminosity possibly does not contribute to the dust heating and should be subtracted when converting into the \sfrprime\ heating the dust in the disk. This fraction is close to the lower limit of the  observed discrepancy and thus might be able to explain  a considerable part of it.

\subsubsection{A different geometry of dust and stars in NGC~4214}

In our modeling we have assumed that NGC~4214 is a scaled-down version of the model spiral galaxy used in \citet{2011A&A...527A.109P}. However the differences in the geometry for the spiral and dwarf galaxies can affect the comparison of the data with the model of \citet{2011A&A...527A.109P}.

The relative scale-heights and scale-lengths of the dust and the stars are derived from a sample of edge-on spiral galaxies \citep{1999A&A...344..868X}. However, the geometry of dwarf galaxies is  known to be different from that of larger spiral galaxies. In particular,  their stellar disks are thicker in comparison to their diameter \citep[e.g.][]{2006ApJS..162...49H}. This is not unexpected because the scale-height  depends on the mass surface density, which determines the gravitational potential in the vertical direction and the velocity dispersion. Both are  generally  similar  in dwarf galaxies and in spiral galaxies, therefore the {\it absolute} value of the vertical scale-height is expected to be similar, and, since they are smaller than spiral galaxies, the ratio between scale-height and scale-length becomes larger. The gas scale-height is even  higher in absolute  terms than in spiral galaxies \citep{2002ASPC..275...57B}. Thus, we would expect higher ratios between the vertical scale-heights and the radial scale-lengths than the one assumed in the model. In NGC~4214  \citet{1999A&A...343...64M} estimates the vertical stellar scale-height  to be about 200\,pc, i.e. about 1/4 to 1/5 of the radial scale-length. This is at least a factor of 3 higher than the ratio assumed in the \citet{2011A&A...527A.109P} model.   However,  the effects of a difference in the ratio between scale-height and scale-length are not expected to be very large as long as the relative scale ratios between stars and dust are the same. 

Another possible, albeit speculative, difference between the model and the real geometry could be a non-central position of the NW and SE SF regions. Since these regions are dominating the young SF activity in the galaxy and are very likely responsible for a large fraction of the UV emission, their possible position above or below the galactic disk would have an important impact on the dust heating, whereas in large galaxies such an asymmetry is unlikely due to the high number of SF regions.

The geometry of the galaxy can also affect the way we are using the attenuation in the disk. The model for attenuation assumes that the star-forming complexes are distributed in a thin disk, with a radially decreasing exponential distribution. If instead NGC~4214 had most of the star formation occurring in its center, then the model for attenuation in the UV may overpredict the attenuation if the high local concentration of stars has the effect of breaking up the dust layer.

%{\bf In summary, there are a number of possible differences between the assumed model geometry and reality. The consequences of them are difficult to quantify without further modeling, but we do not expect any dramatic effects.}

\subsubsection{An extended dust component}

If the dust is more extended compared to the stellar disk than what is assumed in the model, such an extended dust component will be relatively cold and emit more in the submillimeter. By including such a component we would therefore be able to use a higher value of \sfrprime\ for the rest of the dust and decrease
the discrepancy with the observed UV emission. Furthermore, an extended, cold dust component
could  explain the missing flux at the long wavelengths. \citet{2002ApJ...567..221P} have found indications of such an extended dust component with ISO data in dwarf galaxies, however their data lacked the spatial resolution to directly measure its spatial extent. In order to test whether there are indications for an extended dust component,  we derived the radial scale-lengths from SPIRE 250, 350 and 500\mi\ bands. The rather irregular distribution of the dust emission did not allow us to define elliptical isophotes at all galactocentric radii, instead we derived the median values with the isophotal elliptic annuli  defined in the g-band image.  By carrying out a least square-fit to the radial distribution of these median values, we derived scale-lengths of 835 pc, 920 pc and 1067 pc for SPIRE~250, 350 and 500\mi, respectively. The increase of the scale-lengths from shorter to longer wavelengths is expected because of the decrease of the dust temperature towards the outer parts of the galaxy. The longest wavelength (\SPIRE{500}) best represents  the  scale-length of the dust distribution  but most likely still underestimates it. 

The model of \citet{2011A&A...527A.109P} adopts a dust scale-length of $1.406 \times h_{\rm s}$. With \hs{873} this would predict a dust scale-length of 1227\,pc, which is slightly higher than the measured scale-length at \SPIRE{500} band. Thus, although we cannot exclude the presence of an extended  dust component because the dust scale-length can be longer than that at \SPIRE{500} band, we do not find any clear evidence for its presence. 

\subsubsection{Different dust properties}

Until now we have mainly discussed different geometrical scenarios
for explaining the discrepancies we find in the UV. There is however
another possibility, that the submm mass absorption coefficient of
grains, as specified by the \citet{2001ApJ...548..296W}  model and as used in
the radiation transfer model of \citet{2011A&A...527A.109P}, underestimates the true mass
absorption coefficient of the grains in the submm, leading to a lower 
dust surface densities and a lower disk opacity
in the UV for the observed submm brightness of the diffuse disk. This
is a solution that unfortunately is largely degenerate with 
the assumed geometry of stars and dust. Because of this, and the lack of
independent constraints on the existence of such a dust model, we
cannot draw any conclusion on this issue.

\subsubsection{Physical implications}

In the previous subsections  we discussed a few
possibilities that might have caused the discrepancy in the predicted
level of UV emission. Probably the most likely scenario is  that
dynamical effects associated with a recent burst of SF activity
responsible for the central SF complexes has punched holes through
the diffuse dust layer, through which the UV photons leaving the SF
regions can escape the galaxy without interacting with the diffuse
dust. This would be a natural outcome of a non-steady state behaviour
of SF activity in dwarf galaxies, in which we may be observing
NGC~4214 in a post starburst phase, where the star clusters are no
longer fully cocooned by dust, but to the contrary have evolved to
the point where SN and wind activity  have cleared the dust in the
ISM above the location of the starburst.

The existence of different grain properties is also plausible, though
difficult to distinguish from the other scenarios. This would increase
the inferred gas-to-dust mass ratios compared to the radiation transfer model predictions,
bringing it closer to the values expected from the  observed
metallicity of NGC~4214. Furthermore, dust grains with a higher submillimeter emissivity 
and a flatter frequency dependence would  be able to better explain the
dust emission at long wavelengths (850\mi )  where our model 
presently underpredict the observations by 34\%.

%%%%%%%%%%%%%%%%%
%% CONCLUSIONS %%
%%%%%%%%%%%%%%%%%

\section{Summary and conclusions}

We have analysed and modelled the dust SED of the nearby starbursting
dwarf galaxy NGC~4214. Due to its proximity we were able to derive the
dust SED separately for the emission from the two major massive SF
regions and for the diffuse dust. For the first time this analysis
was done from the perspective of a self-consistent radiation transfer calculation
constrained by the observed SEDs of spatially separated components on
resolved maps. In
making predictions for the UV/optical-FIR/submm SEDs of these
components this  analysis quantitatively takes into account the
level and
colour of UV/optical radiation fields incident on both pc-sized dusty
structures in the SF complexes and on dust grains distributed on kpc
scales in the diffuse ISM, illuminated by a combination of UV light
escaping from the SF regions and the ambient optical radiation
fields, as constrained by UV/optical photometry of the galaxy. The
overall analysis was done using the model predictions and the
formalism from \citet{2011A&A...527A.109P}, while for the detailed
modelling of the two central SF complexes we used the model of \citet{2008ApJS..176..438G}.
%
%We have analyzed and modeled the dust heating and emission in the nearby, starbursting dwarf galaxy NGC~4214. The proximity of NGC~4214 permits an analysis with a good spatial resolution so that we were able to derive the dust SED separately for the emission from the two major massive SF regions and for the diffuse dust. We applied physical models based on Galactic dust properties that take into account the full radiation transport to the observed data: the model of \citet{2008ApJS..176..438G} for the dust emission from the HII regions and their surrounding PDRs and the model of \citet{2011A&A...527A.109P} for the diffuse dust emission which is based on a realistic distribution of the dust and stars in a disk galaxy, as well as for the modeling of the total integrated UV/optical/FIR emission of the galaxy. 
%
The large amount of ancillary data and results from previous studies allowed us to constrain a major part of the input parameters of the models.

We achieved a good agreement between data and models, both for the diffuse dust emission and the dust in HII+PDR regions.
For the comparison with the dust emission from HII+PDR regions we could constrain from observations basically all input parameters
(metallicity, age of star cluster, external pressure, heating capacity and covering factor) with the exception of the gas column density of the PDR. We achieved satisfactory fits for both SF regions with the exception of the 8 \mi\ data points.
Possible reasons for this discrepancy are that the model assumptions (PAH abundance and destruction) are not 
completely adequate for the case of NGC~4214 or that NGC~4214 has an unusually high emission at 8 \mi\ for its metallicity and radiation 
field, which is supported by other studies \citep{2008ApJ...678..804E}.

We could fit  the diffuse dust SED satisfactorily, but we inferred that the  UV emission was severely underpredicted  with respect to the
observed, deattenuated diffuse UV flux. 
We have discussed different  explanations for this discrepancy (escape of UV emission, geometrical effects, a very extended dust disk
and different dust properties). The most plausible one is that part (30-70\%) 
of the UV radiation that escapes from HII+PDR regions
leaves the galaxy unattenuated and is thus not participating in the heating of the diffuse dust, most likely due to a porous ISM.

We derived a global gas-to-dust mass ratio of $350-390$, close to the value expected from the metallicity of
NGC~4214 ($Z = 0.3-0.4~Z_\odot$) of $375-500$.

In summary, this is the first time a full radiation transfer analysis
constrained by the observed SEDs of spatially separated components
has been done for a dwarf galaxy. Corresponding radiation transfer
studies of resolved star-forming dwarf galaxies but with different
inclinations and evolutionary states to NGC~4214 may allow to
distinguish between the alternative physical scenarios shaping the
panchromatic SEDs of these systems.

\begin{acknowledgements}
We  thank the referee, F. Galliano, whose comments helped to improve the clearness of the manuscript,
 Joerg Fischera for insightful discussions about the ISM  and M\'ed\'eric Boquien for help with the HERSCHEL data. 
This work has been supported by the research projects  AYA2007-67625-C02-02  and AYA2011-24728 from the Spanish Ministerio de Ciencia y Educaci\'on and the Junta de Andaluc\'\i a (Spain) grants FQM108.
IH was supported by a PhD grant from the Spanish Ministerio de Ciencia y Educaci\'on (BES-2008-008108).
Part of this research has been supported by the ERG HER-SFR from the EC.
This work is based on observations with the Instituto de Radioastronom\'ia Milim\'etrica IRAM 30\,m.
This research has made use of the NASA/IPAC Extragalactic Database (NED), which is operated by the Jet Propulsion Laboratory, California Institute of Technology, under contract with the National Aeronautics and Space Administration. We also acknowledge the usage of the HyperLeda database (http://leda.univ-lyon1.fr). This research made use of Montage, funded by the National Aeronautics and Space Administration's Earth Science Technology Office, Computational Technnologies Project, under Cooperative Agreement Number NCC5-626 between NASA and the California Institute of Technology. The code is maintained by the NASA/IPAC Infrared Science Archive.
\end{acknowledgements}

%%%%%%%%%%%%%%%%%%%%%%
%%%% BIBLIOGRAPHY %%%%
%%%%%%%%%%%%%%%%%%%%%%

\bibliographystyle{aa}
\bibliography{Bibliography}

\end{document}